\documentclass[12pt,a4paper]{article}

\usepackage[T1]{fontenc}
\usepackage[utf8]{inputenc}
\usepackage{lmodern}
\usepackage{amssymb}
\usepackage{amsthm}
\usepackage{amsmath}
\usepackage{lineno}
\usepackage{subcaption}
\usepackage{enumitem}  
\usepackage{empheq} 
\usepackage{lineno}
\usepackage{xcolor}
\usepackage{authblk}

\title{Quasi-Sierpinski Structure for Uniform Load Distribution}

\author[1]{Javier Rodríguez-Cuadrado}

\author[2]{Jesús San Martín}

\affil[1,2]{Departamento de Matemática Aplicada a la Ingeniería Industrial, Escuela Técnica Superior de Ingenieros Industriales, Universidad Politécnica de Madrid, 28006, Madrid, Spain}
\affil[1]{Corresponding author. Mail: javier.rodriguez.cuadrado@upm.es}
\date{}

\begin{document}
\maketitle            
\begin{abstract}
Land reclamation methods, indispensable for the proper development of modern coastal cities, are ecologically destructive. We present a fractal structure, similar to a Sierpinski triangle, which solves this problem by resting directly on the seabed thanks to the uniform load distribution we achieve on its base. To obtain this uniform distribution, we show that the supports of the structure must displace vertically following any function of the Takagi class. This causes the vertical deformations of the structure to follow this same class and the horizontal deformations to be related to the Cantor function. The structure works with an unlimited number of combinations of areas of its elements and materials, which gives designers a high degree of constructive flexibility.
\end{abstract}




\section{Introduction}
The high population density in many coastal areas, particularly in the Indo-Pacific, generates a lack of land on which to provide the services and infrastructures of a modern city. This forces governments to expand cities towards the sea; the cases of Dubai, Hong Kong, Netherlands, Mumbai, Singapore and Jakarta, among others, are well known. \cite{sengupta2018building,sengupta2023mapping}.

Land reclamation is not a recent technique; let us remember stilt houses or polders in the Netherlands. The main current techniques for land expansion towards the sea are i) drainage, ii) floating constructions, iii) piling and iv) filling the seabed, which we explain below: 

\begin{itemize}
\item[i)] Drainage: Dikes are traditionally used to close areas of shallow water and then drain the water to create dry land. The classic example is the polders in the Netherlands mentioned above. However, this technique, as indicated, can only be used to build in shallow areas.

\item[ii)] Floating constructions: consist of the arrangement of floating platforms on which the constructions are placed. This technique has multiple advantages such as adaptability to sea level rise and low carbon footprint, but the weakness of their anchorage makes them unsuitable for areas with high winds and waves.

\item[iii)] Piling: This technique consists of introducing piles into the seabed to place constructions on them, such as the Burj Al Arab in Dubai. This technique offers great stability to the structural assembly as it is directly anchored to the ground, but it is expensive as it requires the necessary machinery to drill, and even in some cases where the soil is of low resistance, to reach the bedrock.

\item[iv)] Filling: This method consists of extracting sand, rock and gravel, either from inland (as in the Kansai airport) or from the seabed itself (by dredging, as in the Palm Islands) and placing it on the seabed where the construction is to be built, creating an artificial island. The main problem with this technique is that it destroys the seabed where the dredging takes place and the deposition sludge spreads over an even larger area, further destroying marine life. Note that filling methods obstruct natural ocean currents and hinder the drainage of runoff waters. In fact, the response of water to land reclamation can determine the success, environmental impact and sustainability of a land reclamation project.
\end{itemize}

It follows from the above that the solution must 1) provide stability without the need to drill, 2) avoid filling and associated sludge, 3) allow free water flow, and 4) favor bioengineering.

A solution that meets the indicated requirements was given by the authors in a previous paper \cite{rodriguez2022sierpinski}. This solution consists of a fractal structure \cite{epstein2008stiffness,loong2023modal}, similar to a Sierpinski triangle, which has the property of uniformly distributing the load on its base when a point load is applied on its top vertex. The uniform distribution is the optimal distribution as it requires the minimum soil resistance, allowing the structure to be placed directly on the seafloor without the need for drilling. The structure has several levels, and increasing their number reduces the point pressure on the bottom to the desired value. By placing several of these triangles in parallel, a prism is obtained, and several of these prisms constitute the structure on which to place a platform above sea level. Services can be built on this platform. By its very nature, being a hollowed structure, it allows water to flow through it without obstructing it, thus placing no limitations on currents, tides or runoff water. The structure can be built with a wide variety of materials, which is especially useful for designers. In addition, corals can grow on it and build an artificial reef, thus facilitating the recovery of degraded seabed or amplifying marine life where it already exists.

However, the indicated structure \cite{rodriguez2022sierpinski} has the great limitation that it requires elastic supports whose stiffness is determined by the Takagi curve \cite{takagi1990simple,allaart2012takagi}. This given curve forces that the stiffnesses of the supports have specific values, which leads to the problem of not finding supports with standard stiffness values. The reason why the stiffnesses have to have such specific values is a consequence of the restrictive relationship between the areas of the members of the structure. In order to solve this problem, we generalize the structure given in \cite{rodriguez2022sierpinski}, so that the members can maintain any ratio of area and modulus of elasticity of the material. This implies that in the new structure the stiffnesses are not given by a single curve but by a whole class, which greatly facilitates the constructive execution by finding the stiffnesses that are necessary for the project, including commercial solutions.  The ease of constructive execution is further enhanced in the generalized structure since not only the material of the members can be any construction material, but also the materials can be different member by member, which was not the case in the predecessor structure (see \cite{rodriguez2022sierpinski}). It should be noted that the material above sea level is subject to different wear processes than the submerged material, and the flexibility of execution indicated above economizes its construction. 

This paper is organized as follows. In section 2, for the sake of completeness, we explain the structure introduced by the authors in \cite{rodriguez2022sierpinski}. Subsequently, in this same section, we detail the generalized structure, the we name Universal Quasi-Sierpinski structure. In section 3, we determine the loads and reactions systems associated to the generalized structure, resorting to the Principle of Virtual Work (PVW) \cite{carpinteri2001static,epstein2006fractal,hou2018static} as it was done in \cite{rodriguez2022sierpinski}. By applying this method, we show that the supports of the structure must be elastic and undergo vertical displacements determined by the Takagi class (section 4). Likewise, we determine the vertical and horizontal displacements of the nodes of the structure, which are related to the Takagi class and the Cantor function \cite{fleron1994note}, respectively. These functions have already appeared in other mechanical studies \cite{davey2011analytical,rodriguez2021fractal}. In section 5, we give an example of the calculation of the Universal Quasi-Sierpinski structure, visually relating the deformations of the structure to the fractal functions that govern the displacements of the nodes. Finally, in section 6, we give the conclusions of this work and the discussion of the results.
 
\section{The structure} \label{sec:str}

We seek a structure that optimally distributes on its base a vertical point load applied on its apex. The optimal distribution is the uniform distribution, since it is the distribution that, for a given value of soil resistance, allows the greatest amount of load to be transmitted. A natural way to transform a point load into a uniformly distributed one is to divide the load successively, which suggests a structure that branches its members progressively. Following this idea, we consider a structure that starts from a point, on which the load is applied, from which two members are born and progressively bifurcate until they reach the base. In addition, we consider that at each bifurcation there is a node that allows the rotation between the members that converge to it, turning the structure into a truss. The advantage of considering the structure as a truss is due to the fact that trusses can be installed quickly, are cost-effective and allow long distances to be bridged. In the structural design of trusses it is also considered that their members only support axial forces and that loads are applied on the nodes, a fact that we will use when applying the PVW.

\begin{figure}[h!]
\centering
\includegraphics[width=10cm]{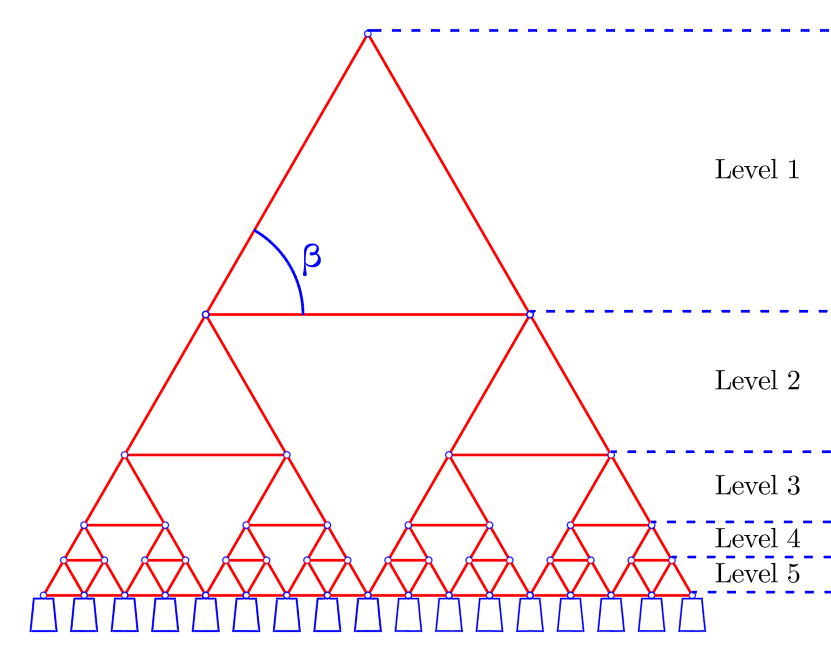}
\caption{Quasi-Sierpinski structure of $N$ levels, corresponding to $N=5$, where $\beta$ is the angle formed by the inclined members with the horizontal.}
\label{fig:qsier}
\end{figure}

In our previous work \cite{rodriguez2022sierpinski}, we presented the $N$-level Quasi-Sierpinski structure (see Fig. \ref{fig:qsier}), a truss capable of transforming a point load into a uniformly distributed load by the progressive bifurcation of its members \footnote{Horizontal members are indispensable so that at each node the correct bifurcation of the load can be achieved}. However, this structure was very limited mechanically, since it required that the cross-sectional area of its members follow a geometric progression of ratio $1/2$ from level to level and that they all be of the same material. In addition, it did not allow horizontal reactions in any of its supports except for its external supports. In the generalized structure, which we call Universal Quasi-Sierpinski structure, we solve all these constraints.

\begin{figure}[h!]
\centering
\includegraphics[width=10cm]{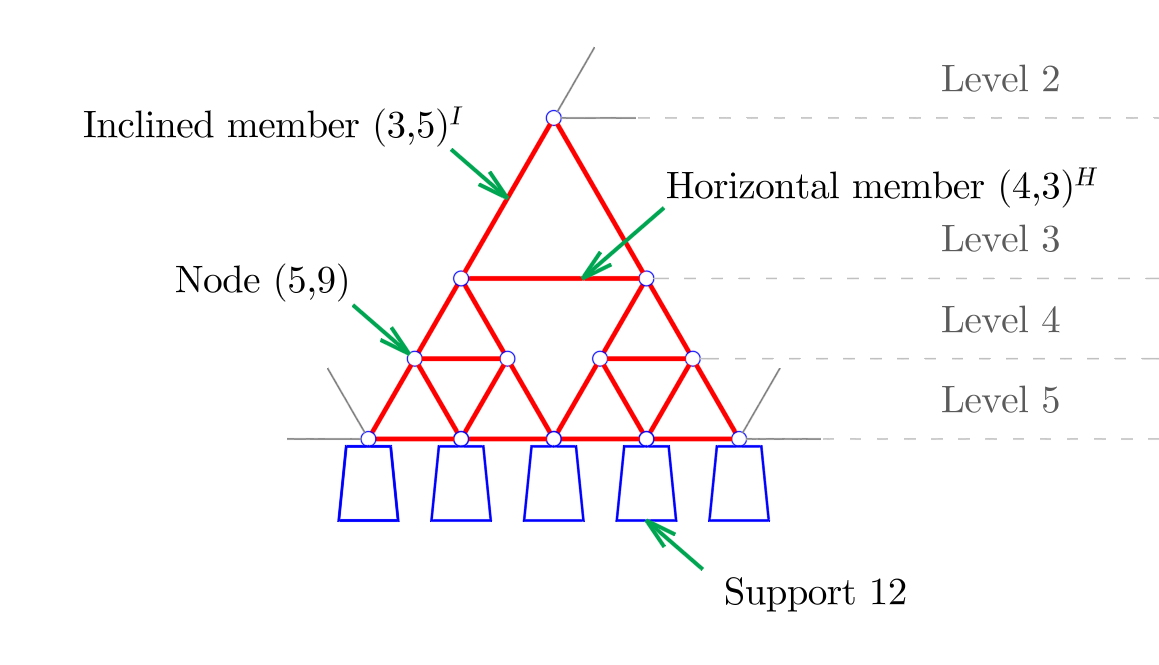}
\caption{Identification of certain nodes, inclined members (I), horizontal members (H) and supports of the Universal Quasi-Sierpinski structure of $N$ levels, corresponding to $N=5$.}
\label{fig:id}
\end{figure}

\begin{figure}[h!]
\centering
\includegraphics[width=10cm]{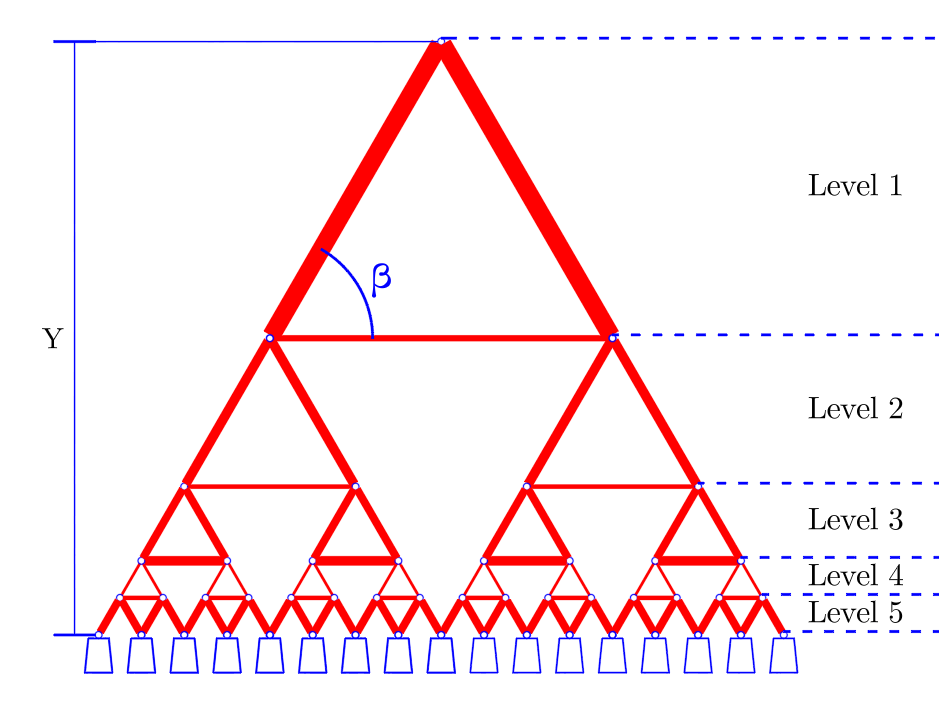}
\caption{Universal Quasi-Sierpinski structure of $N$ levels, corresponding to $N=5$, where $Y$ is the height of the structure.}
\label{fig:Theta}
\end{figure}

The Universal Quasi-Sierpinski structure (see Fig. \ref{fig:Theta}), like its predecessor, comprises $N$ levels and is formed by nodes, inclined members that form an angle $\beta$ with the horizontal, $0 \leq \beta \leq \pi/2$, horizontal members, and supports, which we will identify as follows:

\begin{itemize}
\item Nodes: Identification is performed based on level $n, n=1,2,\ldots,N+1$, and within each level, according to the natural number $t$ that indicates its ordinal position, from left to right, with $t=1,2,\ldots,2^{n-1}$ for $n=1,2,\ldots,N$ and $t=1,2,\ldots,2^{n-2}+1$ for $n=N+1$. In particular, we denote by $(n,t)$ the node of level $n$ that is in the $t$-th position. The nodes of level $N+1$ are the nodes of the supports. Fig. \ref{fig:id} shows the identification of node $(5,9)$ as an example.

\item Inclined members: Identification is performed based on level $n, n=1,2,\ldots,N$, and within each level, according to the natural number $p$ that indicates its ordinal position, from left to right, with $p=1,2,\ldots,2^{n}$. In particular, we denote by $(n,p)^I$ the inclined member of level $n$ that is in the $p$-th position. Fig. \ref{fig:id} shows the identification of the inclined member $(3,5)^I$ as an example.

\item Horizontal members: Identification is performed based on level $n, n=2,3,\ldots,N+1$, and within each level, according to the natural number $q$ that indicates its ordinal position, from left to right, with $q=1,2,\ldots,2^{n-2}$. In particular, we denote by $(n,q)^H$ the horizontal member of level $n$ that is in the $q$-th position. Fig. \ref{fig:id} shows the identification of the horizontal member $(4,3)^H$ as an example.

\item Supports: Identification is performed according to the natural number $i$, which represents the ordinal position of the support, from left to right, with $i=1,2,\ldots,2^{N-1}+1$. Fig. \ref{fig:id} shows the identification of support $12$ as an example.
\end{itemize}

We begin the study with the assumption, which will be proven false in Section \ref{sec:loa}, that the Universal Quasi-Sierpinski structure can have fixed horizontal and vertical supports. Therefore, according to this assumption, the supports do not allow vertical or horizontal displacement. For this reason, the horizontal members in level $N+1$ will not be compressed or stretched, and therefore will not bear any force. This means that, on a practical level, we can disregard them, so the horizontal members range from level $2$ to level $N$.

The geometry of the Universal Quasi-Sierpinski structure, based on the geometry of the Quasi-Sierpinski structure \cite{rodriguez2022sierpinski}, consists of (see Fig. \ref{fig:Theta}):

\begin{itemize}
\item Node $(1,1)$ connects two inclined members on level $1$. Any node on level $n$, $n=2,3,\ldots,N$, connects an inclined member on level $n-1$ with three members on level $n$: one horizontal and two inclined. Nodes $(N+1,1)$ and $(N+1, 2^{N-1}+1)$ connect an inclined member on level $N$ with a support, while the remaining nodes on level $N+1$ connect two inclined members on level $N$ with a support.

\item The length of the inclined members of level $n, n=1,2,\ldots,N-1$, is $\frac{Y}{s 2^n}$, and of level $n=N$ is $\frac{Y}{s 2^{n-1}}$, where $Y$ is the height of the structure and $s = \sin(\beta)$.

\item The length of the horizontal members of level $n, n=2,3,\ldots,N$, is $\frac{c Y}{s 2^{n-2}}$, where $c = \cos(\beta)$.
\end{itemize}

Once the structure has been defined geometrically, we assign it the following mechanical parameters, which generalize those of its predecessor:

\begin{itemize}
\item We denote by $A^I$ and $E^I$ the cross-sectional area and Young's modulus, respectively, of the inclined members of the first level. We denote by $A^H$ and $E^H$ the cross-sectional area and Young's modulus, respectively, of the horizontal members on the second level (note that the horizontal members begin at this level).

\item We denote by $\rho_n^I$ the ratio between the product of the cross-sectional area and Young's modulus of the inclined members of level $n$ and the product of the cross-sectional area and Young's modulus of the inclined members of the first level, with $n=1,2,\ldots,N, \rho_n^I>0$ and $\rho_1^I=1$. Thus, the product of the area and Young's modulus of the inclined members of level $n$ is given by $\rho_n^I A^I E^I$. Finally, we denote by $P_N^I=\{\rho_1^I,\rho_2^I,\ldots,\rho_N^I\}$ the set of ratios of the inclined members.

\item We denote by $\rho_n^H$ the ratio between the product of the cross-sectional area and Young's modulus of the horizontal members of level $n$ and the product of the cross-sectional area and Young's modulus of the horizontal members of the second level, with $n=2,3,\ldots,N, \rho_n^H>0$ and $\rho_2^H=1$. Thus, the product of the area and Young's modulus of the inclined members of level $n$ is given by $\rho_n^H A^H E^H$. Finally, we denote by $P_N^H=\{\rho_2^H,\rho_2^H,\ldots,\rho_N^H\}$ the set of ratios of the horizontal members.
\end{itemize}

Finally, we denote by Q the Universal Quasi-Sierpinski structure that depends on the geometric parameters $N,\beta,Y$ and mechanical parameters $A^I,E^I,A^H,E^H,P_N^I,P_N^H$ described above (see Fig. \ref{fig:Theta}). Note that this structure generalizes the structure defined in \cite{rodriguez2022sierpinski}, since we remove the restriction that fixes the ratios between the areas of the inclined and horizontal members of the different levels. As a result of this generalization, both results and applications are radically more general than those of the structure discussed in \cite{rodriguez2022sierpinski} since, instead of that one solution, we obtain an infinite set of solutions.

\section{Loads and reactions systems and compatibility of deformations} \label{sec:loa}

Structure Q aims to distribute the load uniformly among its supports when applying a downward vertical load of value $F$ on its node $(1,1)$. In particular, given that structure Q has $2^{N-1}+1$ supports, supports $1$ and $2^{N-1}+1$ must bear a downward vertical load of $F/2^N$, and supports $2,3,\ldots,2^{N-1}$ must bear a downward vertical load of $F/2^{N-1}$, since they cover twice the area of the previous ones. Therefore, the reactions on each support will be the opposite forces to those just mentioned.

Note that the required load and reaction system does not specify what the horizontal reactions on the supports should be. However, the geometry of the structure and the required vertical reactions imply that supports $1$ and $2^{N-1}+1$ generate a horizontal reaction of value $F c/2^N s$ to the right and left, respectively, while the remaining supports do not generate a horizontal reaction (see the proof in \ref{sec:appA}).

This system of loads and reactions generates a force of $\frac{-F}{2^n s}$ on the inclined members in level $n, n=1,2,\ldots,N$ and a force of $\frac{F c}{2^{n-1} s}$ on the horizontal members in level $n, n=2,3,\ldots,N$, where the negative sign indicates compression and the positive sign indicates tension. This load system complies with the static equilibrium equations, but the structure is hyperstatic, so it must comply with as many deformation compatibility equations as indicated by its degree of hyperstaticity. In particular, a structure Q with $N$ levels has $3 \cdot 2^{N-1}$ nodes, $5 \cdot 2^{N-1}-3$ members, and $2 \cdot (2^{N-1}+1)$ reactions. Therefore, its degree of hyperstaticity is: $5 \cdot 2^{N-1}-3 + 2 \cdot (2^{N-1}+1) - 2(3 \cdot 2^{N-1}) = 2^{N-1}-1$. Thus, the load system must satisfy $2^{N-1}-1$ linearly independent deformation compatibility equations. These equations are given by the Principle of Virtual Work (PVW) \cite{rodriguez2022sierpinski} which, when applied to structure Q, results in

\begin{equation}
\sum_{nodes} U^V\xi^R = \sum_{members} f^V\Delta l^R,
\label{eq:ptv}
\end{equation}

where:
\begin{itemize}
\item $f^V$ is the force in a member according to the virtual system of loads and reactions (from now on, virtual force),
\item $\Delta l^R$ is the variation in length of a member according to the real system of loads and reactions, with $\Delta l^R = \frac{f^R l}{A E}$, where $f^R$ is the force in that member according to that real system (from now on, real force), $l$ is the length of that member, $A$ is the area of its cross section, and $E$ is the Young's modulus of the material of that member.
\item $U^V$ is the external force (i.e., the load or reaction) on a node according to the virtual system of loads and reactions, and
\item $\xi^R$ is the displacement of the node according to the real system of loads and reactions (from now on, real displacement).
\end{itemize}

In order for both sides of the PVW equation to be dimensionless, we divide both sides of Eq. \ref{eq:ptv} by the height $Y$ of the structure, obtaining

\begin{equation}
\sum_{nodes} U^V\frac{\xi^R}{Y} = \sum_{members} f^V \frac{f^R l}{Y A E}.
\label{eq:ptvn}
\end{equation} 

Note that $\frac{\xi^R}{Y}$ is the real displacement of the node per unit height of the structure. In this paper, we work with three different types of displacements per unit height:

\begin{itemize}
\item We denote by $\varepsilon_{n,t}$ the vertical displacement per unit height of node $(n,t)$, where $\varepsilon_{n,t}>0$ ($<0$) indicates upward (downward) displacement.

\item We denote by $\mu_{n,t}$ the horizontal displacement per unit height of node $(n,t)$, where $\mu_{n,t}>0$ ($<0$) indicates displacement to the right (left).
 
\item We denote by $\delta_i$ the vertical displacement per unit height of support $i$, where $\delta_i>0$ ($<0$) indicates upward (downward) displacement. Note that the displacement of the support is equal to the displacement of the node of that support, i.e., $\delta_i = \varepsilon_{N+1,i}$.
\end{itemize}

\subsection{Virtual loads and reactions system VS-SU} \label{sec:vssu}

\begin{figure}[h!]
\centering
\includegraphics[width=10cm]{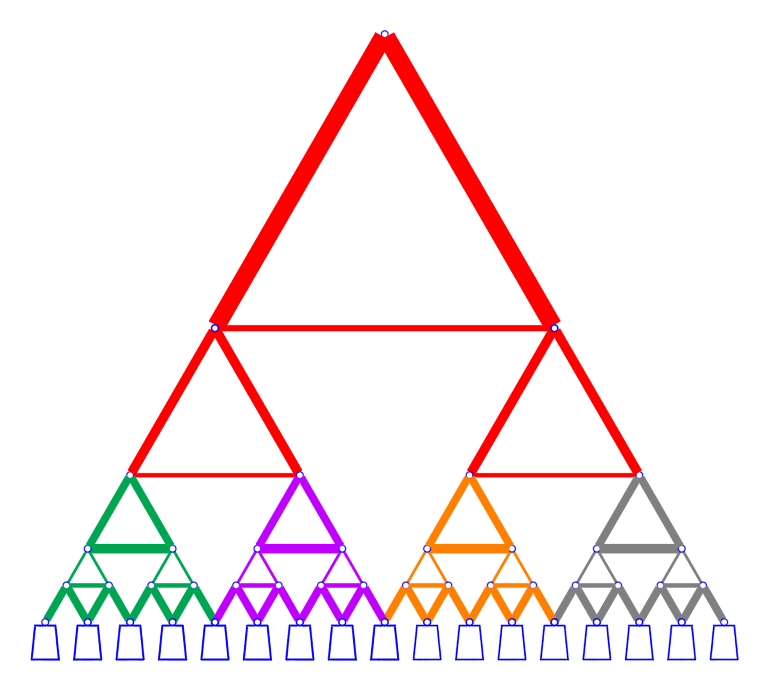}
\caption{Structures Q of $N'$ levels (first in green, second in purple, third in orange and fourth in gray) in a structure Q of $N$ levels, corresponding to $N'=3$ and $N=5$.}
\label{fig:uesm}
\end{figure}

In this subsection, we define linearly independent virtual load and reaction systems that will give rise to linearly independent PVW equations (Eq. \ref{eq:ptvn}). To define these virtual systems, we note that each node of level $n, n=1,2,\ldots,N-1$, can be interpreted as node $(1,1)$ of a structure Q of $N'=N-n+1$ levels, $N'=2,3,\ldots,N$ (see Fig. \ref{fig:uesm}). In particular, there are $2^{N-N'}$ structures Q of $N'$ levels each in a structure Q of $N$ levels, making a total of $2^{N-1}-1$ structures. Therefore, by associating a virtual system with each structure Q of $N'$ levels, we obtain the $2^{N-1}-1$ linearly independent deformation compatibility equations indicated above.

\begin{figure}[h!]
\centering
\includegraphics[width=10cm]{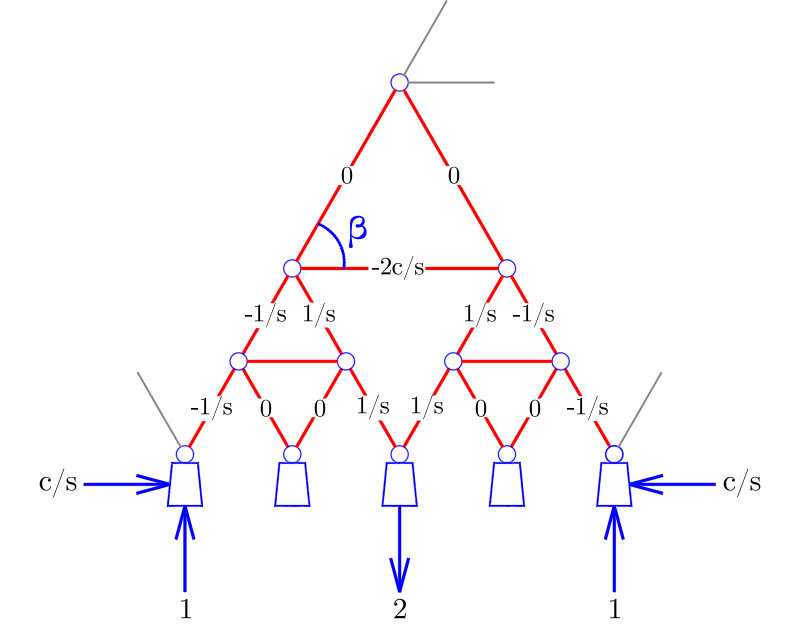}
\caption{System VS-SU and forces generated in the members of the third structure Q of $N'$ levels in a structure Q of $N$ levels, corresponding to $N'=3$ and $N=5$ (see Fig. \ref{fig:uesm}), where $c=\cos(\beta)$ and $s=\sin(\beta)$.}
\label{fig:vlss}
\end{figure}

To work with these structures of $N'$ levels, we denote by VS-SU the virtual system of loads and reactions associated with the ($u+1)$-th $N'$-level structure Q, with $u=0,1,\ldots,2^{N-N'}-1$, which consists of applying the following loads to its supports (see Fig. \ref{fig:vlss}): 
\begin{itemize}
\item[i)] Support $u \cdot 2^{N'-1}+1$: an upward vertical load of value $1$ and a horizontal load to the right of value $c/s$,
\item[ii)] Support $u \cdot 2^{N'-1}+2^{N'-2}+1$: a downward vertical load of value $2$,
\item[iii)] Support $(u+1)2^{N'-1}+1$: an upward vertical load of value $1$ and a horizontal load to the left of value $c/s$.
\end{itemize}
	
On the other hand, the system VS-SU also generates the following forces in the members of the $N'$-level structure:
\begin{itemize}
\item Inclined members. In level $n=N-N'+2, N-N'+3,\ldots,N$, the inclined members $2^{n+N'-N}u+1$ and $2^{n+N'-N}(u+1)$ bear a force of value $-1/s$, and the members $2^{n+N'-N-1}(2u+1)$ and $2^{n+N'-N-1}(2u+1)+1$ bear a force of value $1/s$. The rest of the members bear zero force. 

\item Horizontal members. In level $n=N-N'+2$, the horizontal member $u+1$ bears a force of value $-2c/s$. The rest of the members bear zero force.
\end{itemize}

\subsection{Supports with vertical springs} \label{sec:svsp}

Let us recall that we assumed that the supports were fixed, and therefore the displacements of the support nodes are zero. Consequently, the left side of Eq. \ref{eq:ptvn} is zero since the virtual loads of the system VS-SU are only applied to the supports. On the other hand, when completing the right side of Eq. \ref{eq:ptvn}, there are the same number of inclined members for which the virtual force is $1/s$ and for which the virtual force is the opposite, $-1/s$. These members counteract each other pair by pair because they have the same length, real force, area, and Young's modulus. Thus, on the right side of Eq. \ref{eq:ptvn}, only the following remains:

\begin{gather} 
-\frac{2c}{s} \frac{F c}{2^{(N-N'+2)-1}s} \frac{Y c}{2^{(N-N'+2)-2}s} \frac{1}{Y \rho^H_{N-N'+2} A^H E^H} = \nonumber \\
 = -\frac{F c^3}{2 A^H E^H s^3 } \frac{2}{4^{N-N'} \rho^H_{N-N'+2}}
\label{eq:rptvn}
\end{gather}
\newpage
Note that expression \ref{eq:rptvn} is only zero if $F=0$ \footnote{Other conditions such as $c=0$ could also occur, but they do not make sense from an engineering point of view.}; consequently, the deformation compatibility equations would not be satisfied. Given the structure and the reactions required to obtain a uniform distribution, we must reject our initial assumption of having fixed horizontal and vertical supports and instead allow vertical displacements at the supports\footnote{Horizontal displacements could also be allowed, but the horizontal reaction on all supports except two is zero, so there would be no way to generate the required displacements.}. Therefore, we will consider that each support $i$ can be displaced vertically by an amount $\delta_i$ per unit of height. In this case, given the system VS-SU defined in Section \ref{sec:vssu}, the left side of Eq. \ref{eq:ptvn} is equal to\footnote{The sign is given as a function of the agreement between the reaction in the virtual and real systems: a positive sign indicates the same direction; a negative sign indicates the opposite direction}:

\begin{equation} 
\delta_{u \cdot 2^{N'-1}+1} -2\delta_{u \cdot 2^{N'-1}+2^{N'-2}+1} + \delta_{(u+1)2^{N'-1}+1}
\label{eq:lptvn}
\end{equation}

Taking $m=N-N'$ in expressions \ref{eq:rptvn} and \ref{eq:lptvn} and equaling them, we have that Eq. \ref{eq:ptvn} according to the system VS-SU in each structure Q of $m-N$ levels generates the following system of equations:

\begin{gather}
\delta_{u \cdot 2^{N-m-1}+1} -2\delta_{u \cdot 2^{N-m-1}+2^{N-m-2}+1} + \delta_{(u+1)2^{N-m-1}+1} = -\Omega^H \frac{2}{4^m \rho^H_{m+2}}, \label{eq:ptvf}\\
m=0,1,\ldots,N-2, \nonumber\\
u = 0,1,\ldots,2^m-1, \nonumber
\end{gather}
where $\Omega^H = \frac{F c^3}{2 A^H E^H s^3}$ is a dimensionless parameter given by the characteristics of the structure and the applied load.

In total, there are $2^{N-1}-1$ equations in the system of equations \ref{eq:ptvf}. On the other hand, there are $2^{N-1}+1$ supports with their respective displacements, so two more equations are needed for the system to have a unique solution. These are the boundary equations, which we obtain by setting the value of the displacements per unit height $d_1$ and $d_2$ of any two supports $z_1$ and $z_2$:
\begin{gather}
\delta_{z_1} = d_1,\\
\delta_{z_2} = d_2, 
\label{eq:}
\end{gather}
where $z_1,z_2 \in \{1,2,\ldots,2^{N-1}+1\}$.

\subsection{Virtual loads and reactions system VS-NV} \label{sec:vsnv}

Once the equations determining the vertical displacement of the supports have been established, it is important to know the displacement of the nodes of the structure in order to quantify the deformation of the structural assembly. The vertical and horizontal displacement of each node determines the final position of the nodes, and thus the shape of the structure Q when it is loaded and generates uniform distribution.

\begin{figure}[h!]
\centering
\includegraphics[width=10cm]{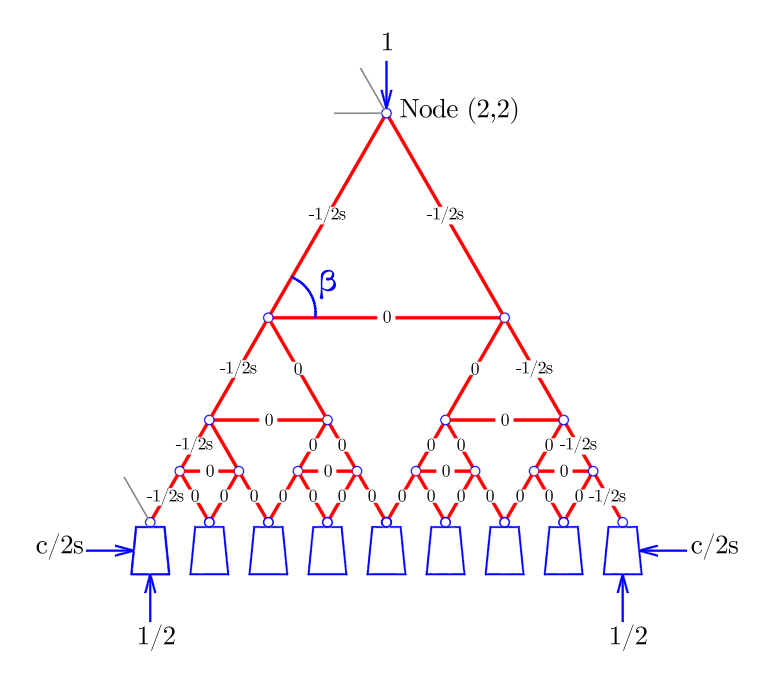}
\caption{System VS-NV and forces generated in the members of the structure Q of $N$ levels to compute the vertical displacement per unit of structure height $\varepsilon_{n,t}$, corresponding to $n=2$, $t=2$ and $N=5$, where $c=\cos(\beta)$ and $s=\sin(\beta)$.}
\label{fig:vdls}
\end{figure}

First, we calculate the vertical displacement per unit height $\varepsilon_{n,t}$ of node $(n,t)$. For nodes in level $N+1$, which are the support nodes, we have $\varepsilon_{N+1,t} = \delta_t$, $t = 1,2,\ldots,2^{N-1}+1$.  For $1 \leq n \leq N$, we establish a virtual system of loads and reactions, called the system VS-NV, which consists of i) on node $(n,t)$, applying a downward vertical load of value $1$, ii) on support $(t-1) 2^{N-n}+1$, applying a vertical upward load of value $\frac{1}{2}$ and a horizontal load to the right of value $\frac{c}{2s}$, and iii) on support $t2^{N-n}+1$, applying an upward vertical load of value $\frac{1}{2}$ and a horizontal load to the left of value $\frac{c}{2s}$ (see Fig. \ref{fig:vdls}). Therefore, the left side of Eq. \ref{eq:ptvn} given the system VS-NV is

\begin{equation}
-\varepsilon_{n,t} + \frac{1}{2} \delta_{(t-1) 2^{N-n}+1} + \frac{1}{2} \delta_{t 2^{N-n}+1}.
\label{eq:lptvvd}
\end{equation}

On the other hand, the system VS-NV only generates forces in the following inclined members (see Fig. \ref{fig:vdls}): the members $(k,2^k(t-1))^I$ and $(k,2^k t)^I$, with $k=n, n+1, \ldots, N$, bear a force of value $\frac{-1}{2s}$. Therefore, the right side of Eq. \ref{eq:ptvn} becomes

\begin{equation}
2 \frac{-1}{2s} \frac{-F}{2^{N}s} \frac{Y}{2^{N-1}s} \frac{1}{Y \rho^I_{N} A^I E^I} + 
2 \sum_{k=n}^{N-1} \frac{-1}{2s} \frac{-F}{2^ks} \frac{Y}{2^k s} \frac{1}{Y \rho^I_k A^I E ^I} = \Omega^I \left(\frac{2}{4^N \rho^I_N} + \sum_{k=n}^{N-1} \frac{1}{4^k \rho^I_k}\right),
\label{eq:rptvvd}
\end{equation}
where $\Omega^I = \frac{F}{A^I E^I s^3}$ is a dimensionless parameter given by the characteristics of the structure and the applied load.

Considering the left side of Eq. \ref{eq:ptvn} given the system VS-NV (Eq. \ref{eq:lptvvd}) and the right side (Eq. \ref{eq:rptvvd}), we have that the vertical displacement $\varepsilon_{n,t}$ can be expressed as a function of the displacements of the supports:

\begin{equation}
\varepsilon_{n,t} =  \frac{1}{2} \delta_{(t-1) 2^{N-n}+1} + \frac{1}{2} \delta_{t 2^{N-n}+1} - \Omega^I \left(\frac{2}{4^N \rho^I_N} + \sum_{k=n}^{N-1} \frac{1}{4^k \rho^I_k}\right).
\label{eq:ptvvd}
\end{equation}

\subsection{Virtual loads and reactions system VS-NH}

In this subsection, we calculate the horizontal displacement per unit height $\mu_{n,t}$ of node $(n,t)$ in a similar way to that used in Section \ref{sec:vsnv}. For nodes in level $N+1$, such as support nodes, we have that 

\begin{equation}
\mu_{N+1,t} = 0,\quad t = 1,2,\ldots,2^{N-1}+1.
\label{eq:hds}
\end{equation}

\begin{figure}[h!]
\centering
\includegraphics[width=10cm]{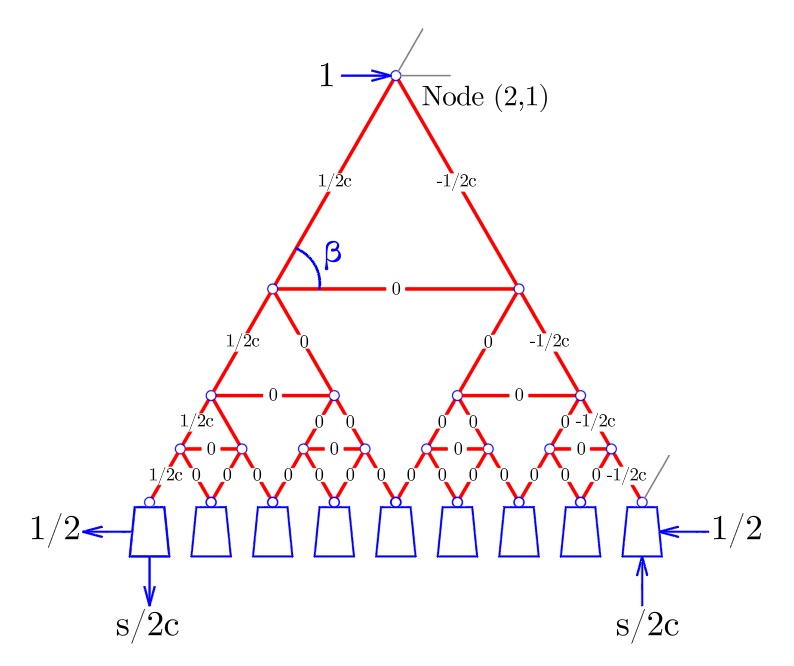}
\caption{System VS-NH and forces generated in the members of the structure Q of $N$ levels to compute the horizontal displacement per unit of structure height $\mu_{n,t}$, corresponding to $n=2$, $t=1$ and $N=5$, where $c=\cos(\beta)$ and $s=\sin(\beta)$.}
\label{fig:hdls}
\end{figure}

For $n = 1,2,\ldots,N$ and $t = 1,2,\ldots,2^{N-1}+1$, we establish a virtual system of loads and reactions, called the system VS-NH, which consists of i) on node $(n,t)$, applying a horizontal load to the right with a value of $1$, ii) on support $(t-1) 2^{N-n}+1$, applying a downward vertical load of value $\frac{s}{2c}$ and a horizontal load to the left of value $\frac{1}{2}$, and iii) on support $t2^{N-n}+1$, applying a vertical upward load of value $\frac{s}{2c}$ and a horizontal load to the left of value $\frac{1}{2}$ (see Fig. \ref{fig:hdls}). Therefore, the left side of Eq. \ref{eq:ptvn} given the system VS-NH is

\begin{equation}
\mu_{n,t} - \frac{s}{2c} \delta_{(t-1) 2^{N-n}+1} + \frac{s}{2c} \delta_{t 2^{N-n}+1}, \quad n=1,2,\ldots,N.
\label{eq:lptvhd}
\end{equation}

On the other hand, the system VS-NH only generates forces in the following inclined members (see Fig. \ref{fig:hdls}): member $(k,2^k(t-1))^I$ bears a force of value $\frac{1}{2c}$ and member $(k,2^k t)^I$ bears a force of value $\frac{-1}{2c}$, with $k=n, n+1, \ldots, N$. Since, in each level, both members have the same geometric and mechanical characteristics and their force is of equal magnitude but opposite sign, then the right-hand side of Eq. \ref{eq:ptvn} given the system VS-NH is zero. Therefore, considering the left side of Eq. \ref{eq:ptvn} given this system (Eq. \ref{eq:lptvhd}), we have that the vertical displacement $\mu_{n,t}$ can be expressed as a function of the displacements of the supports:

\begin{equation}
\mu_{n,t} = \frac{s}{2c} \delta_{(t-1) 2^{N-n}+1} - \frac{s}{2c} \delta_{t 2^{N-n}+1}, \quad n=1,2,\ldots,N.
\label{eq:ptvhd}
\end{equation}

In the following section, we will give an expression for the displacements of the supports, and with it, for the displacements of the nodes according to Eqs. \ref{eq:ptvvd} and \ref{eq:ptvhd}.
\section{Results} \label{sec:res}

\subsection{Displacements of the supports and the Takagi class} \label{sec:dsup}

We seek an individualized explicit expression for vertical displacements instead of the recursive expression given by Eq. \ref{eq:ptvf}. To do this, we write the vertical displacement $\delta_i$ of support $i$ as follows: 

\begin{equation}
\delta_i = \Omega^H g\left(\frac{i-1}{2^{N-1}}\right)+\frac{d_2-\Omega^H g\left(\frac{z_2-1}{2^{N-1}}\right)-\left(d_1-\Omega^H g\left(\frac{z_1-1}{2^{N-1}}\right)\right)}{z_2-z_1}(i-z_1)+d_1-\Omega^H g\left(\frac{z_1-1}{2^{N-1}}\right),
\label{eq:del}
\end{equation}
with $g : [0,1] \rightarrow \mathbb{R}$ being a function to be determined. Note that, according to Eq. \ref{eq:del}, for $i = z_1$ we have $\delta_{z_1} = d_1$ and for $i = z_2$ we have $\delta_{z_2} = d_2$. According to Eq. \ref{eq:del}, and choosing the subscript $i$ to match the subscripts of the displacements of the supports in Eq. \ref{eq:ptvf}, we have:

\begin{align}
\delta_{u \cdot 2^{N-m-1}+1} &= \Omega^H g\left(\frac{u}{2^m}\right)+\nonumber\\
&+\frac{d_2-\Omega^H g\left(\frac{z_2-1}{2^{N-1}}\right)-\left(d_1-\Omega^H g\left(\frac{z_1-1}{2^{N-1}}\right)\right)}{z_2-z_1}(u \cdot 2^{N-m-1}+1-z_1)+\nonumber\\
&+d_1-\Omega^H g\left(\frac{z_1-1}{2^{N-1}}\right), \label{eq:dela}\\
\delta_{u \cdot 2^{N-m-1}+2^{N-m-2}+1} &= \Omega^H g\left(\frac{2u+1}{2^{m+1}}\right)+\nonumber\\
&+\frac{d_2-\Omega^H g\left(\frac{z_2-1}{2^{N-1}}\right)-\left(d_1-\Omega^H g\left(\frac{z_1-1}{2^{N-1}}\right)\right)}{z_2-z_1}(u \cdot 2^{N-m-1}+2^{N-m-2}+1-z_1)+\nonumber\\
&+d_1-\Omega^H g\left(\frac{z_1-1}{2^{N-1}}\right), \label{eq:delb}\\
\delta_{(u+1)2^{N-m-1}+1} &= \Omega^H g\left(\frac{u+1}{2^m}\right)+\nonumber\\
&+\frac{d_2-\Omega^H g\left(\frac{z_2-1}{2^{N-1}}\right)-\left(d_1-\Omega^H g\left(\frac{z_1-1}{2^{N-1}}\right)\right)}{z_2-z_1}((u+1)2^{N-m-1}+1-z_1)+\nonumber\\
&+d_1-\Omega^H g\left(\frac{z_1-1}{2^{N-1}}\right), \label{eq:delc}
\end{align}

Substituting Eqs. \ref{eq:dela}, \ref{eq:delb}, and \ref{eq:delc} into Eq. \ref{eq:ptvf}, we obtain

\begin{gather}
g\left(\frac{2u+1}{2^{m+1}}\right)-\frac{1}{2} \left(g\left(\frac{u}{2^{m}}\right)+g\left(\frac{u+1}{2^{m}}\right)\right) = \frac{1}{4^m \rho^H_{m+2}}, \label{eq:func} \\
m=0,1,\ldots,N-2, \nonumber\\
u = 0,1,\ldots,2^m-1, \nonumber
\end{gather}

The solution of the system of functional equations given by Eq. \ref{eq:func} for $N \rightarrow \infty$ has a continuous and unique solution if and only if $\sum_{m=0}^\infty \lvert \frac{1}{4^m \rho^H_{m+2}} \rvert < \infty$, which is given by \cite{okada1996generalization}:

\begin{equation}
g(x) = g(0) + (g(1)-g(0))L_{1/2}(x) + G(x; P^H_\infty),
\label{eq:gsol}
\end{equation}

where $L_{1/2}$ is a distribution function\footnote{We do not delve further into this function because it is not relevant to this work.}, $P^H_\infty$ is a set that we define in Section \ref{sec:str}, and $G(x; P^H_\infty)$ is the function given by

\begin{equation}
G(x; P^H_\infty) = \sum_{m=0}^\infty \frac{1}{4^m \rho^H_{m+2}} \psi(2^m x),
\label{eq:G}
\end{equation}
where $\psi(2^m x) = \lvert 2x-2 \left\lfloor x+1/2 \right\rfloor \rvert$ is twice the nearest integer distance function, where $\lfloor \cdot \rfloor$ is the floor function. The function $G$ given by Eq. \ref{eq:G} represents the entire Takagi class \cite{hata1984takagi}.

The solution to the system of Eq. \ref{eq:func} given by Eq. \ref{eq:gsol} is particularly valid for any finite $N$. In this case, to calculate the displacements $\delta_i$, $i=1,2,\ldots,2^{N-1}+1$, of the supports according to Eq. \ref{eq:del}, we evaluate the function $g$, and consequently the function $G$ (see Eq. \ref{eq:gsol}), for values $x$ of the form $\frac{i-1}{2^{N-1}}$. Note that, for $m \geq N-1$, we have that $\psi\left(2^m \frac{i-1}{2^{N-1}} \right) = \psi\left(\frac{i-1}{2^{N-m-1}} \right) = 0$ since $\frac{i-1}{2^{N-m-1}}$ is an integer. Therefore, we have that Eq. \ref{eq:G} becomes

\begin{gather}
G\left(\frac{i-1}{2^{N-1}}; P^H_\infty \right) = \sum_{m=0}^{N-2} \frac{1}{4^m \rho^H_{m+2}} \psi\left(\frac{i-1}{2^{N-m-1}} \right) = G \left(\frac{i-1}{2^{N-1}}; P^H_N \right), \label{eq:eqG}\\
i=1,2,\ldots,2^{N-1}+1, \nonumber
\end{gather}
which is a finite sum and does not need to be checked for convergence. Furthermore, it makes engineering sense since the set $P^H_\infty = \{\rho^H_2, \rho^H_3,\ldots,\rho^H_N, \rho^H_{N+1},\ldots\}$ only has the elements $\rho^H_2$ to $\rho^H_N$ defined, corresponding to the existing levels of the structure. 

Defining:
\begin{gather}
\Lambda_1 = d_1 - \Omega^H G\left(\frac{z_1-1}{2^{N-1}}; P^H_N \right), \label{eq:Del1}\\
\Lambda_2 = d_2 - \Omega^H G\left(\frac{z_2-1}{2^{N-1}}; P^H_N \right), \label{eq:Del2}\\
\chi = \frac{\Lambda_2-\Lambda_1}{z_2-z_1}, \label{eq:chi}
\end{gather}
and, substituting Eqs. \ref{eq:gsol}, \ref{eq:G}, \ref{eq:Del1}, \ref{eq:Del2}, and \ref{eq:chi} into Eq. \ref{eq:del}, we write the displacement $\delta_i$ compactly as

\begin{equation}
\delta_i = \Omega^H G\left(\frac{i-1}{2^{N-1}}; P^H_N \right) + \chi(i-z_1) + \Lambda_1, \quad i=1,2,\ldots,2^{N-1}+1.
\label{eq:delf}
\end{equation}

Next, we will pay attention to two fundamental facts:
\begin{itemize}
\item[i)] Note that, according to Eq. \ref{eq:delf}, the displacements of the supports do not depend on the mechanical characteristics of the inclined members. This means that there are structures Q with different areas and Young's moduli in their inclined members that must have the same displacement of their supports in order to obtain a uniform load distribution. Therefore, these characteristics are not key to generating such uniform load distribution.
\item[ii)] On the contrary, analyzing Eq. \ref{eq:delf}, the mechanical characteristics of the horizontal members, the preset displacements $d_1$ and $d_2$, and the supports $z_1$ and $z_2$ for which these displacements are set, are relevant for obtaining a uniform load distribution. In particular, given that a vertical downward load is applied to each support, the displacements per unit height of all supports must be negative. Therefore, supports $z_1$ and $z_2$ and displacements $d_1$ and $d_2$ must be chosen such that $\delta_i < 0$ for all $i = 1,2,\ldots,2^{N-1}+1$.
\end{itemize} 

\subsubsection{Displacements as an infinite variety of fractal functions}

In this subsection, we prove that the displacements given by Eq. \ref{eq:delf} correspond to values that can be taken by an infinite number of fractal functions, including smooth functions, which is a very important improvement over the Quasi-Sierpinski structure presented in \cite{rodriguez2022sierpinski}. According to Eq. \ref{eq:eqG}, Eq. \ref{eq:delf} can be rewritten as:
\begin{equation}
\delta_i = \Omega^H G\left(\frac{i-1}{2^{N-1}}; P^H_\infty \right) + \chi(i-z_1) + \Lambda_1, \quad i=1,2,\ldots,2^{N-1}+1,
\label{eq:delff}
\end{equation}
where, even if the structure has $N$ levels, we can extend the set $P^H_N$ with any fictitious ratios $\rho^H_{N+1}, \rho^H_{N+2}, \ldots$, provided that $\sum_{m=0}^\infty \lvert \frac{1}{4^m \rho^H_{m+2}} \rvert < \infty$, to obtain the set $P^H_\infty = \{\rho^H_2, \rho^H_3,\ldots,\rho^H_N, \rho^H_{N+1},\rho^H_{N+2},\ldots\}$. The objective is to prove that the displacements $\delta_i$ of the supports of the same structure Q with $N$ levels can be obtained from an infinite number of fractal functions $G$ of the Takagi class, as many as the different sets $P^H_\infty$ obtained by extending the same set $P^H_N$ determined by the mechanical characteristics of the structure.

\begin{figure}[h]
\centering
	\begin{subfigure}[b]{6.75cm} 
	\centering
	\includegraphics[width=\textwidth]{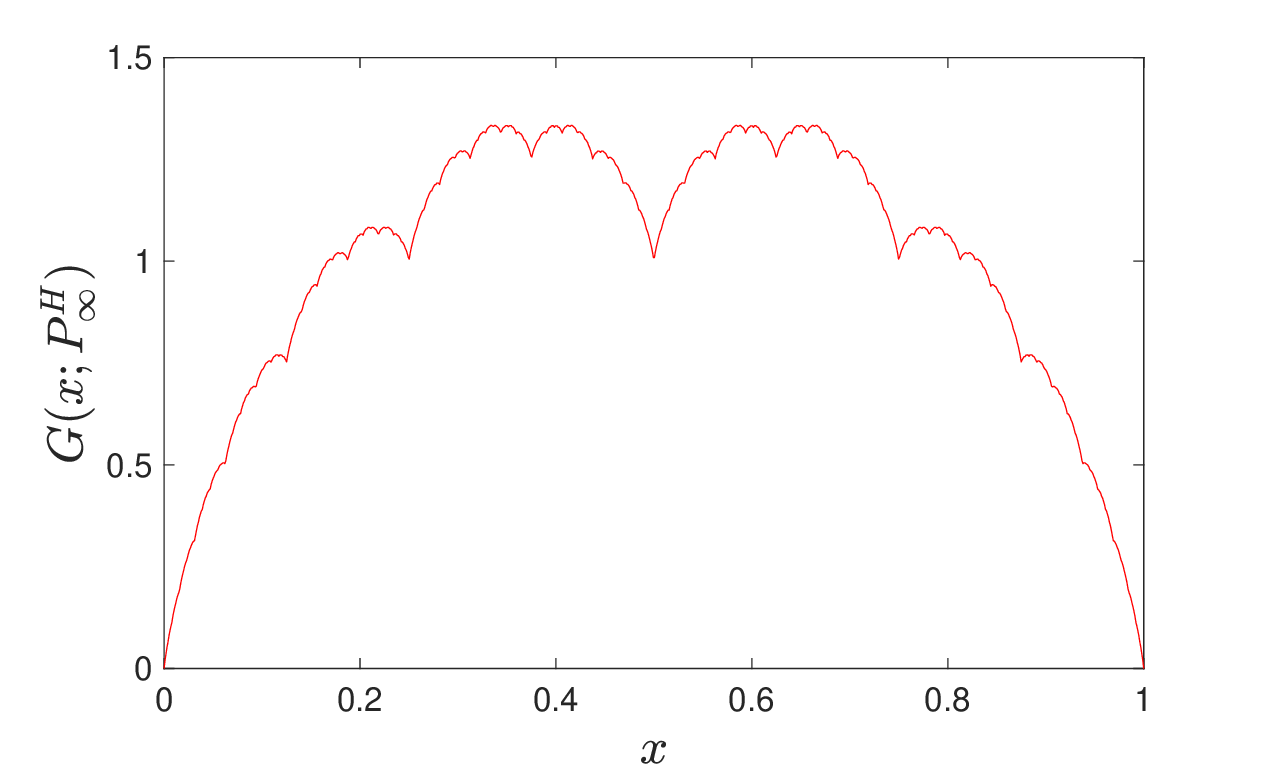}
	\caption{}
	\label{fig:takagi}
	\end{subfigure}
	\hfill
	\begin{subfigure}[b]{6.75cm} 
	\centering
	\includegraphics[width=\textwidth]{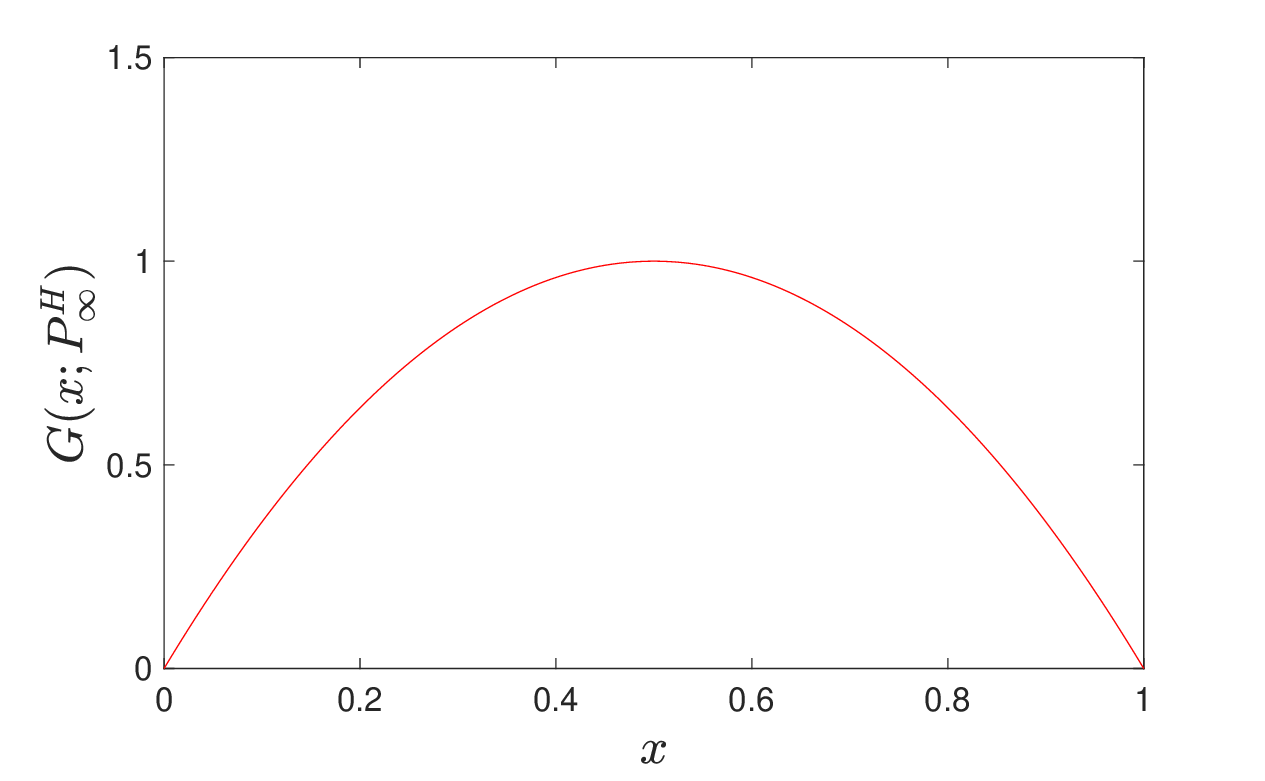}
	\caption{}
	\label{fig:parabola}
	\end{subfigure}
	\caption{Examples of functions $G\left(x; P^H_\infty \right)$ from the Takagi class. In particular, (a) Double of the Takagi function, for $P^H_\infty = \{\left(\frac{1}{2}\right)^k\}_{k=0}^\infty$ and (b) Parabola (a smooth function), for $P^H_\infty = \{1\}_{k=0}^\infty$.}
\end{figure}

For the sake of clarity, we present two examples of functions of the Takagi class:
\begin{itemize}
\item Let $P^H_\infty = \{\left(\frac{1}{2}\right)^k\}_{k=0}^\infty$, then we have that $G(x; P^H_\infty)$ is twice the Takagi curve $T(x)$ (see Fig. \ref{fig:takagi}):
\begin{equation*}
G\left(x; P^H_\infty \right) = \sum_{m=0}^{\infty} \frac{1}{2^m} \psi\left(2^m x\right) = 2T(x)
\end{equation*}

\item Let $P^H_\infty = \{1\}_{k=0}^\infty$, then we have that $G(x; P^H_\infty)$ is a parabola (see Fig. \ref{fig:parabola}):
\begin{equation*}
G\left(x; P^H_\infty \right) = \sum_{m=0}^{\infty} \frac{1}{4^m} \psi\left(2^m x\right) = 4x(1-x),
\end{equation*}
which is a smooth function ($C^\infty$).
\end{itemize}

In general, if the ratios follow a geometric progression with parameter $r$, that is, if $P^H_\infty = \{r^k\}_{k=0}^\infty$, then $G (x; P^H_\infty)$ is twice the Takagi-Landsberg curve with parameter $\frac{1}{4r}$ \cite{landsberg1908differentiierbarkeit}.

Once the relationship between the vertical displacements of the supports and Takagi class has been proved, we will define a fractal function that collects these displacements. We denote by $f_\delta : [0,1] \rightarrow \mathbb{R}$ the function that returns the vertical displacement $\delta_i$ of each support according to its horizontal coordinate $x$. Given that for $i=1$ we have that the value for which we evaluate $G$ in Eq. \ref{eq:delff} is $0$ and for $i=2^{N-1}+1$ the value is $1$, we adopt the criterion that the coordinate $x$ of support $1$ is $0$ and of support $i=2^{N-1}+1$ is $1$, assigning the coordinates of the remaining supports based on this criterion. Therefore, the function $f_\delta$ is given by

\begin{equation}
f_\delta(x; P^H_\infty) = \Omega^H G\left(x; P^H_\infty \right) + \chi(2^{N-1}x+1-z_1) + \Lambda_1.
\label{eq:fdel}
\end{equation}

According to Eq. \ref{eq:delff}, we have that $\delta_i = f_\delta(\frac{i-1}{2^{N-1}}; P^H_\infty)$ for $i = 1,2,\ldots,2^{N-1}+1$. Note that this equality holds for any ratios $\rho^H_{N+1}, \rho^H_{N+2}, \ldots$ that we have chosen to extend the set $P^H_N$, provided that $\sum_{m=0}^\infty \lvert \frac{1}{4^m \rho^H_{m+2}} \rvert < \infty$. On the other hand, according to Eq. \ref{eq:fdel}, $f_\delta$ is an affine transformation of the function $G$, which is a fractal, and therefore $f_\delta$ is a fractal. We therefore deduce that the values of the displacements $\delta_i$ are common to an infinity of fractal functions $f_\delta (x; P^H_\infty)$. This fact gives the structure Q extraordinary constructive flexibility, since the displacements can follow an infinity of functions to obtain a uniform load distribution, not just the Takagi curve.

\subsection{Stiffness of the supports} \label{sec:ssup}
To obtain a uniform load distribution on the base of structure Q, we must ensure that the supports are displaced per unit height in accordance with Eq. \ref{eq:delf}. One way to achieve this is by using elastic supports with a specific stiffness. This stiffness, denoted by $k_i$ for support $i$, $i = 1,2,\ldots,2^{N-1}+1$, is obtained according to Hooke's law and Eq. \ref{eq:delf} as
 
\begin{equation}
k_{i}=\left\{\begin{array}{lc}\displaystyle
\frac{F}{-2^{N} Y \left( \Omega^H G\left(\frac{i-1}{2^{N-1}}; P^H_N \right) + \chi(i-z_1) + \Lambda_1\right)} & \text { if } i=1,2^{N-1}+1 \\[5mm] \displaystyle
\frac{F}{-2^{N-1} Y \left( \Omega^H G\left(\frac{i-1}{2^{N-1}}; P^H_N \right) + \chi(i-z_1) + \Lambda_1\right)} \quad & \text { if } i=2, \ldots, 2^{N-1}
\end{array}\right..
 \label{eq:rig}
\end{equation}

\subsection{Vertical displacements of the nodes} \label{sec:vdis}

In this subsection, we prove that the nodes are vertically displaced following a sum of Takagi-type functions. To do this, we must take into account that the vertical displacement per unit height of the structure $\varepsilon_{n,t}$ of node $(n,t)$ is given as a function of the displacements of the supports (Eq. \ref{eq:ptvvd}). If we substitute the displacement of the supports (Eq. \ref{eq:delf}) into Eq. \ref{eq:ptvvd}, we obtain:

\begin{align}
\varepsilon_{n,t} &= \frac{\Omega^H}{2} \left( G\left(\frac{t-1}{2^{n-1}}; P^H_N\right) + G\left(\frac{t}{2^{n-1}}; P^H_N\right) \right) + \nonumber\\
&+ \chi \left( 2^{N-n-1}(2t-1)+1-z_1\right) + \Lambda_{z_1} -  \Omega^I \left(\frac{2}{4^N \rho^I_N} + \sum_{k=n}^{N-1} \frac{1}{4^k \rho^I_k}\right).
\label{eq:vdnf}
\end{align}

Once again, the nodes are displaced following a sum of Takagi class functions determined by the set $P^H_N$. In this case, the displacement value is also affected by the mechanical characteristics of the inclined members. According to Eq. \ref{eq:eqG}, Eq. \ref{eq:vdnf} can be rewritten as

\begin{align}
\varepsilon_{n,t} &= \frac{\Omega^H}{2} \left( G\left(\frac{t-1}{2^{n-1}}; P^H_\infty\right) + G\left(\frac{t}{2^{n-1}}; P^H_\infty\right) \right) + \nonumber\\
&+ \chi \left( 2^{N-n-1}(2t-1)+1-z_1\right) + \Lambda_{z_1} -  \Omega^I \left(\frac{2}{4^N \rho^I_N} + \sum_{k=n}^{N-1} \frac{1}{4^k \rho^I_k}\right),
\label{eq:vdnff}
\end{align}
expression that we will use a few lines below.

Once the relationship between the vertical displacements of the nodes and the Takagi class has been proved, we will define a fractal function that collects these displacements. We denote by $f_\varepsilon: \{1,2,\ldots,N+1\}\times[0,1] \rightarrow \mathbb{R}$ the function that returns the vertical displacement $\varepsilon_{n,t}$ of each node $(n,t)$ according to its horizontal coordinate $x$. Under the coordinate criteria adopted in Section \ref{sec:dsup}, we define $f_\varepsilon$ as

\begin{align}
f_\varepsilon (n,x; P^H_\infty) &= \frac{\Omega^H}{2} \left( G\left(x-\frac{1}{2^{n}}; P^H_\infty\right) + G\left(x+\frac{1}{2^{n}}; P^H_\infty\right) \right) + \nonumber\\
&+ \chi \left( 2^{N-1}x+1-z_1\right) + \Lambda_{z_1} -  \Omega^I \left(\frac{2}{4^N \rho^I_N} + \sum_{k=n}^{N-1} \frac{1}{4^k \rho^I_k}\right).
\label{eq:feps}
\end{align}

According to Eq. \ref{eq:vdnff}, we have that $f_\varepsilon (n, \frac{2t-1}{2^n}; P^H_\infty) = \varepsilon_{n,t}$ for $t = 1,2,\ldots,2^{n-1}+1$, $n=1,2,...,N+1$. Note that this equality holds for any ratios $\rho^H_{N+1}, \rho^H_{N+2}, \ldots$ chosen to extend the set $P^H_N$ and thus obtain $P^H_\infty$, provided that $\sum_{m=0}^\infty \lvert \frac{1}{4^m \rho^H_{m+2}} \rvert < \infty$. Furthermore, according to Eq. \ref{eq:feps}, $f_\varepsilon$ is an affine transformation of functions $G$, which are fractals. We therefore deduce that the values of the displacements $\varepsilon_{n,t}$ are common to an infinity of functions $f_\varepsilon (n,x; P^H_\infty)$ (which are affine transformations of fractals), as many as possible sets $P^H_\infty$.

\subsection{Horizontal displacements of the nodes} \label{sec:hdis}

In this subsection, we prove that the nodes move horizontally following an affine transformation of Takagi class functions, which in some cases will result in functions related to the Cantor function. To do this, we must take into account that the horizontal displacement per unit height of the structure $\mu_{n,t}$ of node $(n,t), t=1,2,\ldots,2^{N-1}+1, n=1,2,\ldots,N$, is given as a function of the displacements of the supports by Eq. \ref{eq:ptvhd}. If we substitute the displacement of the supports (Eq. \ref{eq:delf}) into Eq. \ref{eq:ptvhd}, we obtain

\begin{equation}
\mu_{n,t} = \frac{\Omega^H s}{2c} \left( G\left(\frac{t-1}{2^{n-1}}; P^H_N\right) - G\left(\frac{t}{2^{n-1}}; P^H_N\right) \right) - \chi \frac{2^{N-n-1}s}{c}.
\label{eq:hdnf}
\end{equation}

Once again, the nodes are displaced following an affine transformation of Takagi class functions determined by the set $P^H_N$. In this case, moreover, the displacement value is not affected by the mechanical characteristics of the inclined members. According to Eq. \ref{eq:eqG}, Eq. \ref{eq:hdnf} can be rewritten as

\begin{equation}
\mu_{n,t} = \frac{\Omega^H s}{2c} \left( G\left(\frac{t-1}{2^{n-1}}; P^H_\infty\right) - G\left(\frac{t}{2^{n-1}}; P^H_\infty\right) \right) - \chi \frac{2^{N-n-1}s}{c}.
\label{eq:hdnff}
\end{equation}

According to Eqs. \ref{eq:hds} and \ref{eq:hdnff}, the horizontal displacements can be rewritten as

\begin{align}
\mu_{n,t} &= \frac{\Omega^H s}{2^{n-2}c} \left(2J\left(\frac{2t-1}{2^{n}}; P^H_\infty\right) - \frac{1}{2^{n-1} \rho^H_{n+1}} - \sum_{k=0} ^{n-2} \frac{1}{2^{k+1} \rho^H_{k+2}} \right) - \chi \frac{2^{N-n-1}s}{c},\nonumber\\
& n = 1,2,\ldots,N,\nonumber\\
\mu_{n,t} &= 0, \quad n = N+1
\label{eq:hdalt}
\end{align}
where

\begin{equation}
J\left(x; P^H_\infty\right) = \sum_{k=0}^\infty \frac{\gamma_k(x)}{2^{k+1} \rho^H_{k+2}}
\label{eq:J}
\end{equation}
and $\gamma_k(x)$ are the coefficients of the dyadic expansion of $x$ (see \ref{sec:appB}).

\begin{figure}[h]
\centering
	\begin{subfigure}[b]{6.75cm} 
	\centering
	\includegraphics[width=\textwidth]{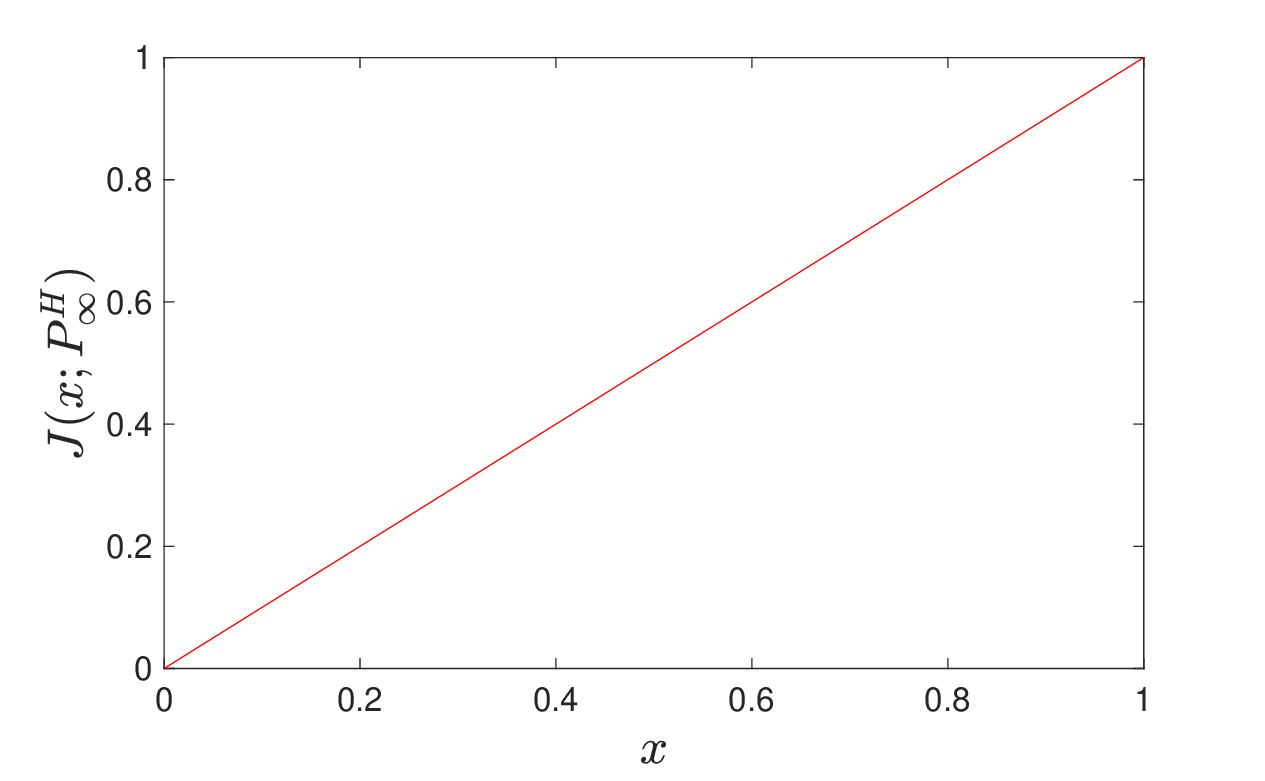}
	\caption{}
	\label{fig:line}
	\end{subfigure}
	\hfill
	\begin{subfigure}[b]{6.75cm} 
	\centering
	\includegraphics[width=\textwidth]{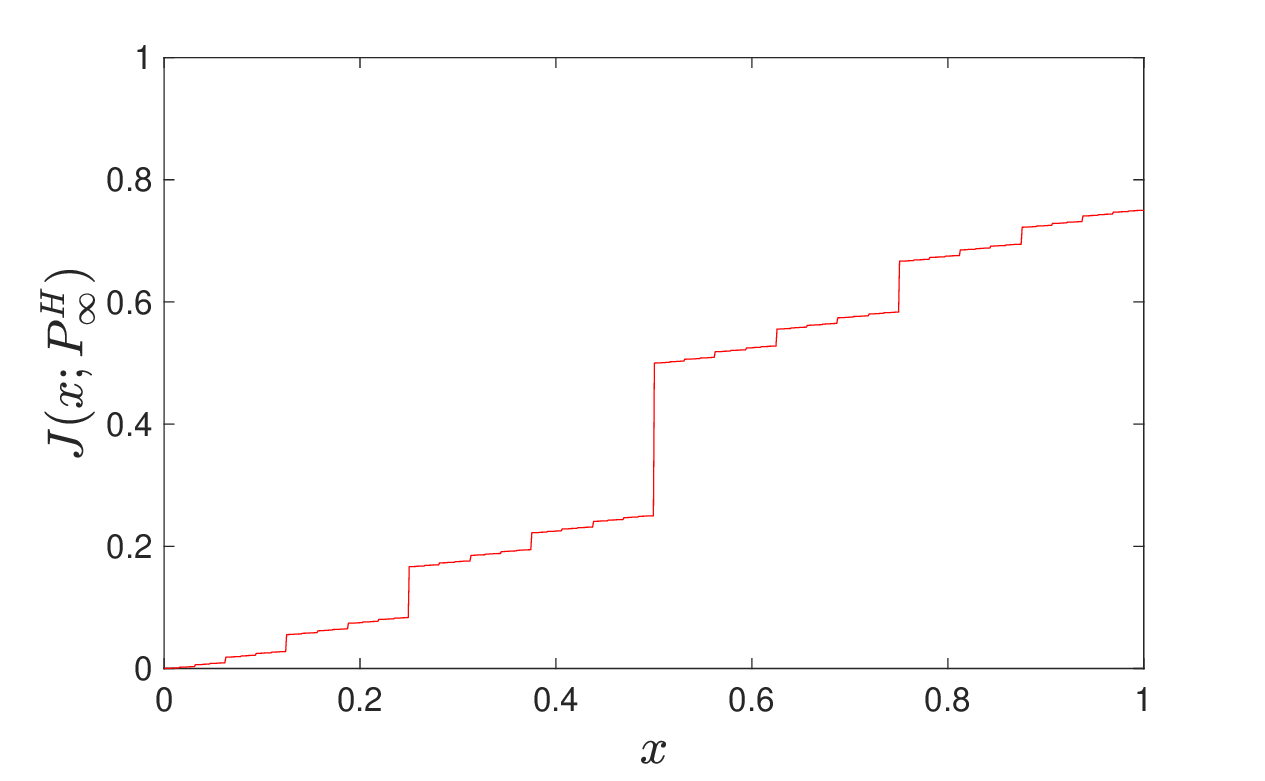}
	\caption{}
	\label{fig:cantor}
	\end{subfigure}
	\caption{Examples of functions $J\left(x; P^H_\infty \right)$. In particular, (a) Identity function, for $P^H_\infty = \{1\}_{k=0}^\infty$ and (b) $\frac{3}{4}$ times the pseudo-inverse of the Cantor function, for $P^H_\infty = \{\left(\frac{3}{2}\right)^k\}_{k=0}^\infty$.}
\end{figure}

For the sake of clarity, and to illustrate the horizontal displacements, we present two examples of functions $J\left(x; P^H_\infty\right)$:
\begin{itemize}
\item Let $P^H_\infty = \{1\}_{k=0}^\infty$, we have that
\begin{equation*}
J(x; P^H_\infty) = \sum_{k=0}^\infty \frac{\gamma_k(x)}{2^{k+1}} = x,
\end{equation*}
so it is the identity function (see Fig. \ref{fig:line}).

\item Let $P^H_\infty = \{\left(\frac{3}{2}\right)^k\}_{k=0}^\infty$, we have that 
\begin{equation*}
J\left(x; P^H_\infty \right) = \frac{3}{4}\sum_{k=0}^{\infty} \frac{2\gamma_k(x)}{3^{k+1}},
\end{equation*}
so it is $\frac{3}{4}$ times the pseudo-inverse Cantor function \cite{mesiarova2009ranks} (see Fig. \ref{fig:cantor}). 
\end{itemize}

In Fig. \ref{fig:cantor}, we see how, for some sets $P^H_\infty$, the function $J\left(x; P^H_\infty \right)$ is fractal. In general, if $P^H_\infty = \{r^k\}_{k=0}^\infty$, then $\frac{r}{2r-1}J\left(x; P^H_\infty \right)$ is the Cantor pseudo-inverse based on the bases $2r$ and $2$ \cite{mesiarova2009ranks}\footnote {In reference \cite{mesiarova2009ranks}, Cantor pseudo-inverses are continuous on the left, while the functions  $\frac{r}{2r-1}J\left(x; P^H_\infty \right)$ are continuous on the right.}. In particular, note that for $P^H_\infty = \{1\}_{k=0}^\infty$, Cantor's pseudoinverse is based on the bases 2 and 2, so that the identity function is effectively obtained (see Fig. \ref{fig:line}).

Once the relationship between the horizontal displacements of the nodes and the fractals has been proved, we will define a function that collects these displacements. We denote by $f_\mu : \{1,2,\ldots,N+1\}\times[0,1] \rightarrow \mathbb{R}$ the function that returns the horizontal displacement $\mu_{n,t}$ of each node $(n,t)$ according to its horizontal coordinate $x$. Under the coordinate criteria adopted in Section \ref{sec:dsup} above, we define $f_\mu$ as

\begin{align}
f_\mu (n,x; P^H_\infty) &=\frac{\Omega^H s}{2^{n-2}c} \left(2J\left(x; P^H_\infty\right) - \frac{1}{2^{n-1} \rho^H_{n+1}} - \sum_{k=0} ^{n-2} \frac{1}{2^{k+1} \rho^H_{k+2}} \right)+\nonumber\\
&- \chi \frac{2^{N-n-1}s}{c}, \quad n = 1,2,\ldots,N,\\
f_\mu (n,x; P^H_\infty) &= 0, \quad n = N+1.
\label{eq:fmu}
\end{align}

According to Eq. \ref{eq:hdalt}, we have that $f_\mu(n, \frac{2t-1}{2^n}; P^H_\infty) = \mu_{n,t}$ for $t = 1,2,\ldots,2^{n-1}+1$, $n=1,2,...,N+1$. Note that again this equality holds for any ratios $\rho^H_{N+1}, \rho^H_{N+2}, \ldots$ chosen to extend the set $P^H_N$ and thus obtain $P^H_\infty$, provided that $\sum_{m=0}^\infty \lvert \frac{1}{4^m \rho^H_{m+2}} \rvert < \infty$. Furthermore, according to Eq. \ref{eq:fmu}, $f_\mu$ is an affine transformation of the function $J$, which has fractal behavior. Therefore, the values of the displacements $\mu_{n,t}$ are common to an infinity of fractal functions $f_\mu (n,x; P^H_\infty)$, as many as there are possible sets $P^H_\infty$.
\section{Example}

In this section, we perform the study of a structure Q that bears a downward vertical load of $100$ kN on its upper vertex. In particular, we assign the following parameters to the structure: $N=5$, $\beta = 1.1071$, $Y = 16000$ mm, $A^I = 8$ mm\textsuperscript{2}, $E^I = 210$ kN/mm\textsuperscript{2}, $A^H = 0.5$ mm\textsuperscript{2}, $E^I = 210$ kN/mm\textsuperscript{2}, $P^I_N = \{1, 0.5, 0.5, 0.25, 0.25\}$, $P^H_N = \{1, 0.75, 0.5, 0.5\}$. The reader can observe that the value of the areas of both inclined and horizontal members is very low; we do this in order to clearly see the deformed shape of the structure. In this structure, we choose supports $1$ and $17$ to move vertically $-1050$ mm (i.e., downwards), so we take $d_1 = d_2 = -1050/16000$. Finally, we place springs on the supports of the structure whose stiffnesses follow Eq. \ref{eq:rig}. 

\begin{figure}[h!]
\centering
\includegraphics[width=13.2cm]{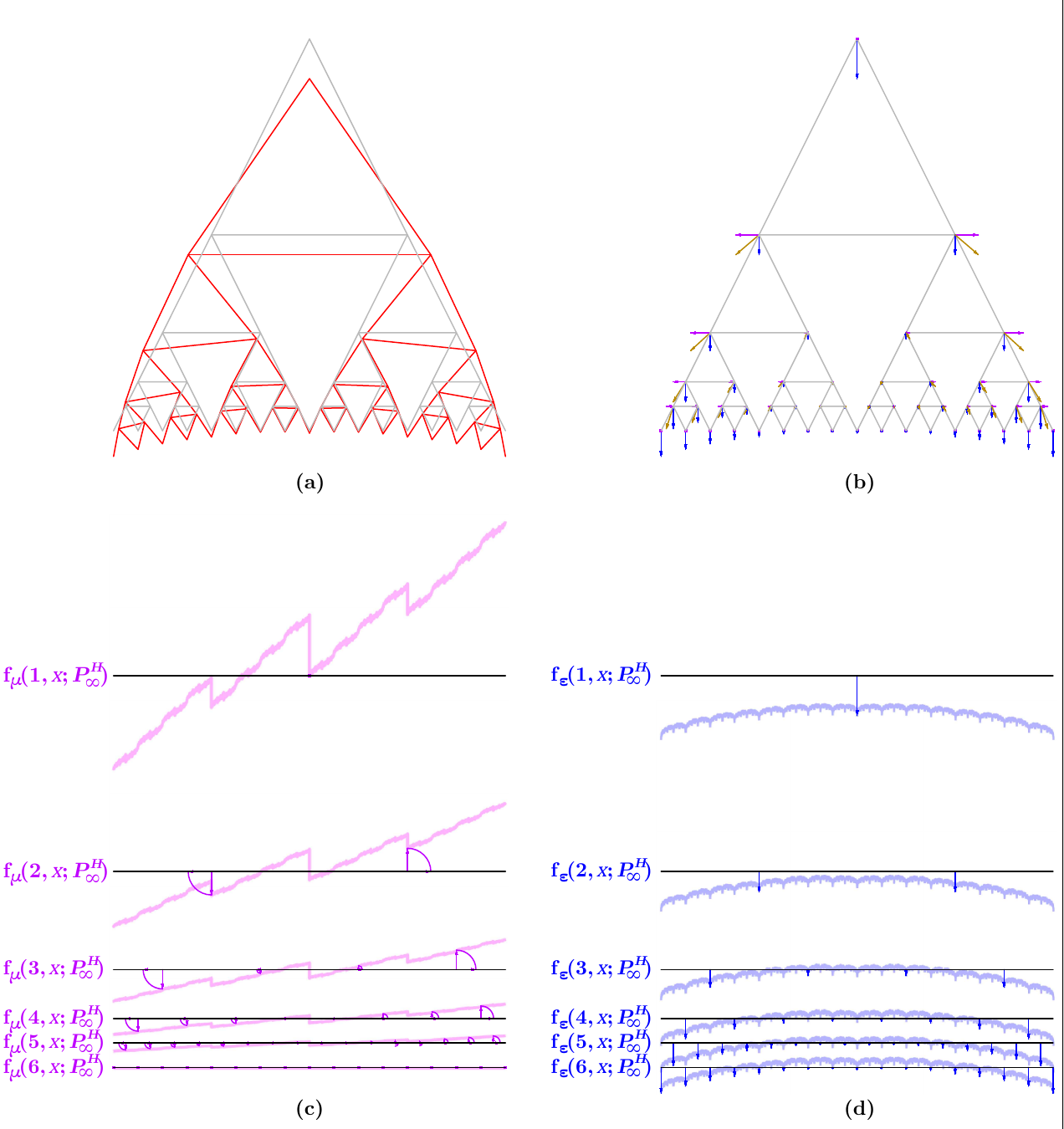}
\caption{(a) Shape of structure Q before (in gray) and after load (in red), (b) Original shape of the structure Q (in gray) and the displacements of its nodes (in brown), decomposed in horizontal (in magenta) and vertical (in blue), (c) Horizontal displacements, their tilt on the vertical direction and functions $f_\mu(1,x; P^H_\infty), f_\mu(2,x; P^H_\infty),\ldots,f_\mu(6,x; P^H_\infty)$ to which they belong, placed on horizontal axes at the same relative distance than the levels of the structure Q, with $P^H_\infty = \{1, 0.75, 0.5, 0.5, 0.3^3, 0.3^4, 0.3^5, \ldots\}$ and (d) Vertical displacements and functions $f_\varepsilon(1,x; P^H_\infty), f_\varepsilon(2,x; P^H_\infty),\ldots,f_\varepsilon(6,x; P^H_\infty)$ to which they belong placed on horizontal axes at the same relative distance than the levels of the structure Q, with $P^H_\infty = \{1, 0.75, 0.5, 0.5, 0.75, 0.75^2, 0.75^3, \ldots\}$.}
\label{fig:example}
\end{figure}

The deformed shape of the structure and the displacements of its nodes were calculated using Robot Structural Analysis Professional \textsuperscript{\textregistered} software. Fig \ref{fig:example}a shows the shape of structure Q before (in gray) and after applying the load (in red). As we can see, the nodes have displaced a certain distance after loading, which is shown in green in Fig \ref{fig:example}b and decomposed into horizontal displacement (in magenta) and vertical displacement (in blue). As mentioned in Section \ref{sec:hdis}, these horizontal displacements belong to an infinite number of functions $f_\mu$. In particular, in Fig \ref{fig:example}c we show the functions $f_\mu(1,x;P^H_\infty), f_\mu(2,x;P^H_\infty),\ldots,f_\mu(6,x;P^H_\infty)$ obtained with the set $P^H_\infty = \{1, 0.75, 0.5, 0.5, 0.3^3, 0.3^4, 0.3^5, \ldots\}$, and the reader can see how the horizontal displacements calculated by the software belong to these functions after being rotated. Also, as mentioned in Section \ref{sec:hdis}, the vertical displacements belong to an infinite number of functions $f_\varepsilon$. In particular, in Fig \ref{fig:example}d we show the functions $f_\varepsilon(1,x; P^H_\infty), f_\varepsilon(2,x;P^H_\infty),\ldots,f_\varepsilon(6,x;P^H_\infty)$  obtained with the set $P^H_\infty = \{1, 0.75, 0.5, 0.5, 0.75, 0.75^2, 0.75^3, \ldots\}$, and the reader can see how the vertical displacements calculated by the software belong to these functions.

\newpage
\section{Conclusions and discussion}

In this paper, we propose a structural design for addressing the problem of construction on collapsible soils. Specifically, we present a prefractal structure that distributes the load uniformly on its base, which is optimal because it ensures the stability of the structure with the lowest bearing capacity of the soil.  
\begin{itemize}
\item In terms of mathematical behavior, the uniform distribution implies:
\begin{itemize}
\item[\tiny \labelitemi] The supports must displace vertically following an affine transformation of a Takagi class function (see Eq. \ref{eq:delff}), which is determined by the mechanical characteristics of the horizontal members.  

\item[\tiny \labelitemi] As a result, nodes that do not belong to the supports move vertically according to an affine transformation of functions of the Takagi class, while they move horizontally according to a class of fractal functions that includes Cantor pseudo-inverses. Both the functions of the Takagi class and those that include Cantor pseudo-inverses are determined by the mechanical properties of the horizontal members.
\end{itemize}

\item In terms of structural design, there are three key variables: the mechanical properties of the horizontal and vertical members, and the displacements at two supports chosen freely by the designer. By adjusting these variables, the designer controls:
\begin{itemize}
\item[\tiny \labelitemi] The required stiffness of the supports, which can be adjusted to meet commercial requirements and thus simplify structural execution. In this case, the required stiffness of the supports is not affected by the mechanical properties of the vertical members.  

\item[\tiny \labelitemi] The vertical and horizontal displacements of nodes that do not belong to the supports. In this case, only the vertical displacements are affected by the mechanical properties of the vertical members.
\end{itemize}

Consequently, the designer has complete freedom to choose the mechanical properties of the vertical members and thus freely vary the final shape of the structure.
\end{itemize}

The flexibility in the choice of the structural design parameters, combined with the wide range of materials suitable for its fabrication, makes Universal Quasi-Sierpinski structure an adaptable and engineering-efficient structural solution.

\newpage
\appendix
\section{Horizontal reactions} \label{sec:appA}

We will prove by induction that supports $2$ to $2^{N-1}$ of an $N$-level structure Q do not generate horizontal reaction when it transforms a point load on its upper vertex into a uniform load on its base. To this end, we propose the induction hypothesis: any structure Q with $N$ levels that supports a downward vertical force of value $F/2$ on its node $(1,1)$ and whose supports generate a uniform vertical reaction (i.e., the vertical reaction is upward on all supports, with a value of $F/2^{N+1}$ on supports $1$ and $2^{N-1}+1$ and of value $F/2^{N}$ on the remaining ones), generates on its supports $1$ and $2^{N-1}+1$ a horizontal reaction of value $F c/ 2^{N+1} s$ to the right and left, respectively, while the remaining supports do not generate a horizontal reaction. The proof by induction consists of the following steps:

\begin{itemize}
\item[i)] First, we will prove that the induction hypothesis is true for a structure Q with two levels. As a result of the uniform distribution of vertical reactions, supports 1 and 3 generate an upward vertical reaction of value F/4, while support 2 generates an upward vertical reaction of value F/2. Applying the laws of static equilibrium at nodes (3,1) and (3,3), we conclude that the horizontal reaction on support 1 is to the right and has a value of $F c/4 s$, and the horizontal reaction on support 3 is to the left and has a value of $F c/4 s$. Applying the laws of static equilibrium to the entire structure, given that the sum of horizontal forces must be equal to zero, we conclude that the horizontal reaction on support $2$ must be zero (see Fig. \ref{fig:apa2}).

\begin{figure}[h!]
\centering
\includegraphics[width=10cm]{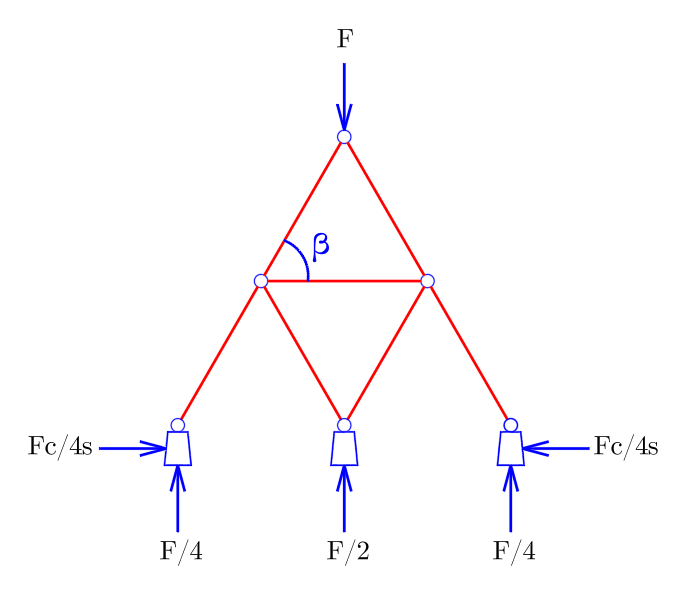}
\caption{Load system in a structure Q of $N$ levels, corresponding to $N=2$, where $c=\cos(\beta)$, $s=\sin(\beta)$ and $F$ is the value of the downward force applied on the node $(1,1)$.}
\label{fig:apa2}
\end{figure}

\item[ii)] Now we assume that the induction hypothesis is true for a structure Q with $N$ levels, and we will prove that, given this assumption, the hypothesis is true for a structure Q with $N+1$ levels. To do this, note that an $N+1$-level structure Q contains two identical $N$-level structures Q, such that the $2^{N-1}+1$ support of the first (in purple, see Fig. \ref{fig:apaN}) is the $1$ support of the second (in green, see Fig. \ref{fig:apaN}). Since the vertical reactions follow a uniform distribution in the $N+1$-level of a structure Q, then support $1$ of the first $N$-level structure Q and support $2^{N-1}+1$ of the second generate an upward vertical reaction of value $F/2^{N+1}$, while supports $2$ to $2^{N-1}$ of the first and second generate an upward vertical reaction of value $F/2^N$. The support $2^{N-1}+1$ of the first structure, coinciding with the support $1$ of the second structure, also generates an upward vertical reaction of value $F/2^N$. Since both $N$-level structures Q are identical and symmetrical with respect to the vertical axis passing through their node $(1,1)$, the analysis can be performed separately for both structures. Therefore, we consider that support $2^{N-1}+1$ of the first one generates an upward vertical reaction of value $F/2^{N+1}$ and support $1$ of the second one generates an upward vertical reaction of value $F/2^{N+1}$ (the sum of both gives the reaction of value $F/2^{N}$ of the structure Q of $N+1$ levels, see Fig. \ref{fig:apaN}).

\begin{figure}[h!]
\centering
\includegraphics[width=10cm]{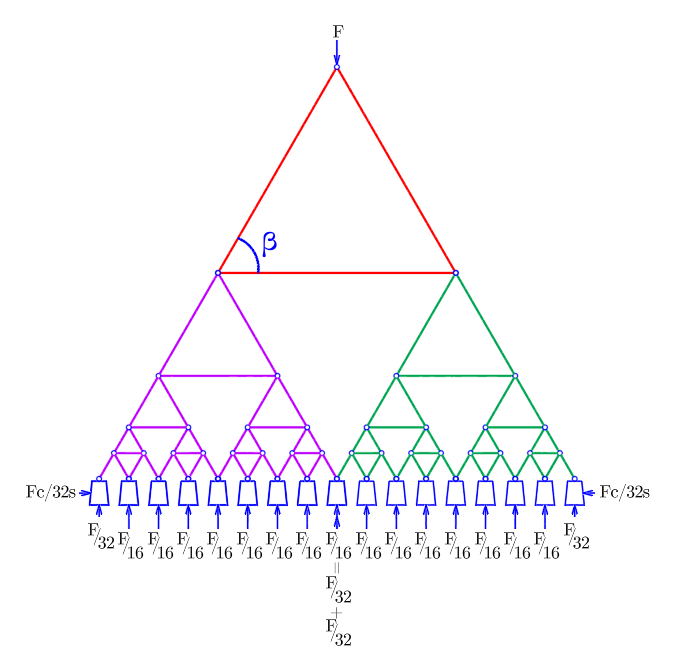}
\caption{Load system in a structure Q of $N+1$ levels and the inclusion of two structures Q of $N$ levels (in purple and green), corresponding to $N=4$, where $c=\cos(\beta)$, $s=\sin(\beta)$ and $F$ is the value of the downward force applied on the node $(1,1)$.}
\label{fig:apaN}
\end{figure}

If we apply the laws of static equilibrium to nodes $(N+1,1)$ and $(N+1, 2^{N-1}+1)$ of both $N$-level structures Q separately, we see that supports $1$ and $2^{N-1}+1$ generate a horizontal reaction of value $F c/ 2^{N+1} s$ to the right and left, respectively. On the other hand, since the sum of vertical forces is equal to zero in each structure, we conclude that the force applied on node $(1,1)$ of each of them must be vertical and downward and have a value of $F/2$. Since the sum of moments is equal to zero, we conclude that there is no horizontal force applied on node $(1,1)$ of each structure. Therefore, applying the induction hypothesis, we conclude that supports $2$ to $2^{N-1}$ of each $N$-level structure Q do not generate horizontal reaction. 

Therefore, supports $2$ to $2^{N-1}$ and $2^{N-1}+2$ to $2^{N}$ of the $N+1$-level structure Q do not generate horizontal reaction, while supports $1$ and $2^N+1$ generate a horizontal reaction of value $F c/2^{N+1} s$ to the right and left, respectively. Since the sum of horizontal forces is zero throughout the structure, we conclude that support $2^{N-1}+1$ also does not generate horizontal reaction.
\end{itemize}

\section{Dyadic expansion} \label{sec:appB}

We show that the horizontal displacements $\mu_{n,t}$ of the nodes given by Eq. \ref{eq:hdnff} can be rewritten in terms of a class of fractal functions that include pseudo-inverses of Cantor functions. According to Eq. \ref{eq:eqG}, we have that the difference between functions $G$ of Eq. \ref{eq:hdnff} can be rewritten as

\begin{equation}
G\left(\frac{t-1}{2^{n-1}}; P^H_\infty\right) - G\left(\frac{t}{2^{n-1}}; P^H_\infty\right) =  \sum_{k=0} ^{n-2} \frac{1}{4^k \rho^H_{k+2}} \left(\psi \left( \frac{t-1}{2^{n-k-1}} \right) - \psi \left( \frac{t}{2^{n-k-1}} \right)\right).
\label{eq:Gdif}
\end{equation}

Note that $\lvert \psi \left( \frac{t-1}{2^{n-k-1}} \right) - \psi \left( \frac{t}{2^{n-k-1}} \right) \rvert = \frac{2}{2^{n-k-1}}$ since the distance between $\frac{t}{2^{n-k-1}}$ and $\frac{t-1}{2^{n-k-1}}$ is $\lvert\frac{t}{2^{n-k-1}}-\frac{t-1}{2^{n-k-1}}\rvert=\frac{1}{2^{n-k-1}}$. On the other hand, the sign of $\psi \left( \frac{t-1}{2^{n-k-1}} \right) - \psi \left( \frac{t}{2^{n-k-1}} \right)$ depends on $t$. If the mean value between $\frac{t}{2^{n-k-1}}$ and $\frac{t-1}{2^{n-k-1}}$, that is, $\frac{1}{2}\left(\frac{t}{2^{n-k-1}}+\frac{t-1}{2^{n-k-1}}\right) = 2^k \frac{2t-1}{2^n}$, is equal to an integer $p$ o or to an integer $p$ plus $\frac{1}{2}$, we have that $\psi \left( \frac{t-1}{2^{n-k-1}} \right) - \psi \left( \frac{t}{2^{n-k-1}} \right)=0$. Furthermore, for $2^k \frac{2t-1}{2^n} \in \left(p, p+\frac{1}{2}\right)$, we have that $\psi \left( \frac{t-1}{2^{n-k-1}} \right) - \psi \left( \frac{t}{2^{n-k-1}} \right)<0$, and for $2^k \frac{2t-1}{2^n} \in \left(p+\frac{1}{2}, p+1\right)$, we have that $\psi \left( \frac{t-1}{2^{n-k-1}} \right) - \psi \left( \frac{t}{2^{n-k-1}} \right)>0$, $p = 0,1,\ldots,2^k-1$. We define

\begin{equation*}
\alpha_{k} \left( \frac{2t-1}{2^{n}} \right) =\left\{\begin{array}{ll}
-1 &\text { if } \frac{2t-1}{2^{n}} \in \left(\frac{p}{2^k}, \frac{p}{2^k}+\frac{1}{2^{k+1}}\right) \\
1 &\text { if } \frac{2t-1}{2^{n}} \in \left(\frac{p}{2^k}+\frac{1}{2^{k+1}}, \frac{p+1}{2^k}\right) \end{array}, p = 0,1,\ldots,2^k-1\right.,
\end{equation*}
and therefore $\psi \left( \frac{t-1}{2^{n-k-1}} \right) - \psi \left( \frac{t}{2^{n-k-1}} \right) = \frac{2 \alpha_{k} \left( \frac{2t-1}{2^{n}} \right)}{2^{n-k-1}}$. Thus, we rewrite Eq. \ref{eq:Gdif} as

\begin{equation}
G\left(\frac{t-1}{2^{n-1}}; P^H_\infty\right) - G\left(\frac{t}{2^{n-1}}; P^H_\infty\right) =  \sum_{k=0} ^{n-2} \frac{1}{4^k \rho^H_{k+2}} \frac{2 \alpha_{k} \left( \frac{2t-1}{2^{n}} \right)}{2^{n-k-1}} = \sum_{k=0} ^{n-2} \frac{\alpha_{k} \left( \frac{2t-1}{2^{n}} \right)}{2^{n+k-2}\rho^H_{k+2}}.
\label{eq:Gdiff}
\end{equation}

Note that the coefficients $\gamma_k$, from $k=0$ to $k=n-1$, of the dyadic expansion of $\frac{2t-1}{2^n}$ are obtained through the coefficients $\alpha_k$ as $\gamma_k \left(\frac{2t-1}{2^n} \right) = \frac{\alpha_k \left(\frac{2t-1}{2^n} \right)+1}{2}$. Therefore:

\begin{equation}
\frac{2t-1}{2^n} = \sum_{k=0}^{n-1} \frac{\gamma_k \left(\frac{2t-1}{2^n} \right)}{2^{k+1}} = \sum_{k=0}^{\infty} \frac{\gamma_k \left(\frac{2t-1}{2^n} \right)}{2^{k+1}},
\end{equation}
where $\gamma_{n-1} \left(\frac{2t-1}{2^n} \right) = 1$ since $\frac{2t-1}{2^n} = \frac{t-1}{2^{n-1}} + \frac{1}{2^n}$, and $\gamma_{k} \left(\frac{2t-1}{2^n} \right) = 0$ for $k \geq n$. Therefore:

\begin{gather}
\sum_{k=0} ^{n-2} \frac{\alpha_{k} \left( \frac{2t-1}{2^{n}} \right)}{2^{n+k-2}\rho^H_{k+2}} = \sum_{k=0} ^{n-2} \frac{\gamma_{k} \left( \frac{2t-1}{2^{n}} \right)}{2^{n+k-3}\rho^H_{k+2}} - \sum_{k=0} ^{n-2} \frac{1}{2^{n+k-2}\rho^H_{k+2}} = \nonumber \\
= \sum_{k=0} ^{n-1} \frac{\gamma_{k} \left( \frac{2t-1}{2^{n}} \right)}{2^{n+k-3}\rho^H_{k+2}} - \frac{\gamma_{n-1} \left( \frac{2t-1}{2^{n}} \right)}{2^{2n-4}\rho^H_{n+1}} - \sum_{k=0} ^{n-2} \frac{1}{2^{n+k-2}\rho^H_{k+2}} = \nonumber \\
= \sum_{k=0} ^{\infty} \frac{\gamma_{k} \left( \frac{2t-1}{2^{n}} \right)}{2^{n+k-3}\rho^H_{k+2}} - \frac{1}{2^{2n-4}\rho^H_{n+1}} - \sum_{k=0} ^{n-2} \frac{1}{2^{n+k-2}\rho^H_{k+2}} \label{eq:alg}
\end{gather}

Consequently, if we take the definition of $J\left(x; P^H_\infty\right)$ given by Eq. \ref{eq:J} and substituting Eq. \ref{eq:alg} into Eq. \ref{eq:Gdiff}, we finally have:

\begin{align*}
G\left(\frac{t-1}{2^{n-1}}; P^H_\infty\right) - G\left(\frac{t}{2^{n-1}}; P^H_\infty\right) &= \frac{1}{2^{n-3}} \biggl( 2J\left(\frac{2t-1}{2^{n}}; P^H_\infty\right)+\\
&-\frac{1}{2^{n-1}\rho^H_{n+1}} - \sum_{k=0} ^{n-2} \frac{1}{2^{k+1}\rho^H_{k+2}} \biggl).
\end{align*}

\bibliographystyle{unsrt} 
\bibliography{ref}

\end{document}